\documentclass{aa}  
\usepackage{graphicx}
\usepackage{txfonts}
\usepackage{multirow}

\begin{document} 

\title{Complex morphology and precession indicators of AGN jets in LoTSS DR2}

   \author{M. A. Horton
          \inst{1\thanks{\email{mah241@cam.ac.uk}},2}
          \and
          M. J. Hardcastle\inst{2}
          \and
          G. K. Miley\inst{3}
          \and
          C. Tasse\inst{4,5}
          \and
          T. Shimwell\inst{3,6}
          }

   \institute{Battcock Centre for Experimental Astrophysics, Cavendish Laboratory, University of Cambridge, JJ Thompson Avenue, Cambridge CB3 0HE, United Kingdom
    \and
    Centre for Astrophysics Research, University of Hertfordshire,
              College Lane, Hatfield, Hertfordshire AL10 9AB, United Kingdom
    \and 
    Leiden Observatory, Leiden University, PO Box 9513,NL-2333CA Leiden, The Netherlands
    \and 
    GEPI \& ORN, Observatoire de Paris, Université PSL, CNRS, 5 Place Jules Janssen, 92190 Meudon, France 
    \and 
    Department of Physics \& Electronics, Rhodes University, PO Box 94, Grahamstown, 6140, South Africa
    \and 
    ASTRON, Netherlands Institute for Radio Astronomy, Oude Hoogeveensedijk 4, 7991 PD, Dwingeloo, The Netherlands
    }

   \date{Received XXXX; accepted XXXX}

  \abstract
{The LOw Frequency ARray Two-metre Sky Survey second data release (LoTSS DR2) covers 27\% of the northern sky and contains around four million radio sources. The development of this catalogue involved a large citizen science project (Radio Galaxy Zoo: LOFAR) with more than 116,000 resolved sources going through visual inspection. We took a subset of sources with flux density above 75 mJy and an angular size of $90''$ or greater, giving a total of $9,985$ sources or $\sim 10\%$ of the visually inspected sources. We classified these by visual inspection in terms of broad source type (e.g., Fanaroff-Riley class I or II, narrow or wide-angle tail, relaxed double), noticeable features (wings, visible jets, banding, filaments etc), environmental features (cluster environment, merger, diffuse emission). Our specific aim was to search for features linked to jet precession, such as a misaligned jet axis, curvature and multiple hotspots. This combination of features and morphology allowed us to detect increasingly fine-grained sub-populations of interesting or unusual sources. We found that $28\%$ of sources showed evidence of one or more precession indicators, which could make them candidates for hosting close binary supermassive black holes. Potential precession signatures occur in sources of all sizes and luminosities in our sample but appear to favour more massive host galaxies. Our work greatly expands the sample size and parameter space of searches for precession signatures in powerful jetted sources. This work also showcases the diversity of large bright radio sources in the LOFAR surveys, whether or not precession indicators are present.
}

\maketitle

\section{Introduction}

Extended radio sources are uniquely powerful astrophysical diagnostics. Because of their large sizes they measure the integrated activity of galaxy nuclei that have been active for up to $10^9$ yr. Their sizes and morphologies have therefore long been used as tools to probe the history of activity in galaxies, the supermassive black holes that power them and the environment in which they are located and as a discriminant between cosmological models \citep[e.g.,][]{miley80}.

Jet morphology plays a critical role in bridging the gap between environment and galaxy dynamics. Early radio surveys of even small numbers of objects, once high-resolution observations were available showed striking variations in structure that resulted in distinct `shapes' or morphological subtypes \citep{Bridle+Perley84}. It was realised at an early stage that some of these morphological classes, such as the head-tail or narrow-angle tail objects, were likely the result of strong interactions with a cluster environment \citep[e.g.,][]{Owen+78,Jones+Owen79} while others, like X-shaped sources and double-double or restarting radio sources, have been interpreted as telling us about activity on scales close to those of the central supermassive black hole \citep[e.g.,][]{merritt02,schoenmakers00b}. However, there are still many sources with ambiguous aetiology, resulting in much debate in the literature, in particular on the topic of environmental versus intrinsic causes of particular observed structures \citep[e.g.,][]{harwood20,hardcastle19}.

Precession in kiloparsec-scale AGN jet paths, which exploits the fact that the jet orientation history is encoded in the large-scale appearance of the source, has long been viewed as a method for identifying close binary supermassive black holes \citep{gower82,krause18}. Curvature in jet paths can be used to place constraints on binary separation where estimates of black hole mass exist \citep{krause18,horton20a}, and key morphological features in double-lobed radio galaxies map directly onto central galactic dynamics, such as precession cone opening angle, precession period, and source age \citep[e.g.,][]{horton20b, horton23}. Although more extreme morphologies are associated with strong geodetic precession from supermassive binaries, many precession-related features could also be produced from warped or misaligned accretion disks \citep[e.g.,][]{fragner10,nixon13,davis20}, the Bardeen-Petterson effect \citep[e.g.,][]{caproni07}, and other causes of jet precession \citep[e.g.,][]{fragile05,stone12}. 

There is currently considerable interest in identifying candidate supermassive binary black holes. Recent population constraints on gravitational wave strain have been produced for supermassive binaries \citep{agazie23}, which will likely become detectable in coming decades \citep[e.g.,][]{buonanno14}. One major challenge will be the separation of supermassive foreground sources \citep[e.g.,][]{babak12,riles13,agazie23b} from a theoretical isotropic gravitational background \citep[e.g.,][]{arzoumanian20,reardon23}. Observational astronomy will be needed to provide a population of candidate galaxies along with good predictions of chirp mass and gravitational wave strain \citep{babak16}, along with some understanding of the closeness, complexity and AGN activity of binary systems \citep{burkespolaor11,pfeifle24}. Therefore it is crucial to understand not only precession morphology of the most likely host galaxies, but whether or not the precession comes from true supermassive binary black holes or environmental factors \citep[e.g.,][]{steinle24,bourne24}. We can only understand the extent to which these precession indicators are affected by real black hole accretion systems through finding a large enough sample population of possible binaries.

Until recently this has been limited due to the lack of sufficient sensitivity and coverage in sky surveys, and by such surveys not being directly focused on precession indicators. That constraint is increasingly being removed for large parts of the sky due to new instruments and deeper surveys. For example, the FIRST survey with the VLA at 1.4 GHz \citep{cheung07,proctor11,yang19} produced a large sample of complex and bent radio sources, including x-shaped, peculiar and possible hybrid sources. Unfortunately the low sensitivity of FIRST, particularly to extended structure, makes it difficult to detect features or be certain of the relationships between source components, which means follow-ups of the same sources are often required to confirm or reject initial classifications.

The second data release of the LOFAR (LOw Frequency ARray) Two-metre Sky Survey (LoTSS) Northern-sky survey (hereafter referred to as LoTSS DR2), is currently the largest radio galaxy survey of the Northern sky, with more than four million detected radio sources \citep{shimwell22}. Because this survey was carried out at a range of 120-168 MHz, with a central frequency of 144 MHz, it is more sensitive to extended radio source morphologies than higher frequency surveys. The optical identification work for this sample, which resulted in more than 3.5 million optical matches overall \citep{hardcastle23} was supported by a citizen science project, Radio Galaxy Zoo: LOFAR, hosted by Zooniverse, where more than 100,000 radio sources were visually inspected by astronomers and citizen scientists. In the process of carrying out this visual inspection we found many previously-unseen objects with complex morphologies. As with earlier surveys, such as FIRST and LoTSS DR1 \cite[e.g.,][]{cheung07,shimwell19,williams19}, the LoTSS survey includes sources with a variety of morphologies. Although we shall comment on these various morphological types, the main purpose of our study is to create a catalogue of radio sources that have morphological signatures of precession, indicative of host galaxies that house supermassive binary black holes.

In this paper we investigate how morphological precession indicators depend on radio power and size (which we use as a proxy for jet power and source age) and whether there are particular types of sources where the environmental distortions dominate or do not dominate (for example, sources where there is evidence of a galaxy cluster which may produce crosswinds or other environmental disturbance). We also aim to investigate whether or not the precession indicators shown in the much smaller 3CRR population seen in \cite{krause18} are representative of the potentially precessing objects found in this much larger sample. 

\section{Sample selection and classification setup}

We make use of the second data release of the LoTSS survey \citep{shimwell22}. LoTSS -- the LOFAR Two-metre Sky Survey -- \citep{shimwell17} is a low-frequency radio survey which aims to cover the whole northern sky at a resolution of $6''$. Relative to other large-area surveys such as NVSS \citep{condon98} and FIRST \citep{becker95} it is both much deeper -- approximately ten times deeper than FIRST -- and also much more sensitive to extended radio structures. Because of this and of its comparatively high resolution LoTSS images allow us both to find optical counterparts for radio galaxies and to characterise their radio morphology.

Optical identifications of LoTSS sources, which use the Legacy and WISE surveys \citep{dey19,meisner18} are provided by a combination of direct radio/optical cross-matching, for relatively simple radio sources, and visual inspection for more complex objects \citep{williams19}. For DR2 the visual inspection has been done through the Radio Galaxy Zoo (LOFAR) citizen science project\footnote{\url{https://www.zooniverse.org/projects/chrismrp/radio-galaxy-zoo-lofar}}, hereafter denoted RGZ(L); the process is described in detail by \cite{hardcastle23}. We selected a subsample of the largest and brightest of these visually inspected sources, requiring that the sources have an optical ID and using an angular size cut of $1.5'$ and a flux cut of $75$ mJy. This was done to keep the subsample manageable while capturing the most bright and resolved galaxies, and was intended to give us a minimum of 15 beam-widths across the source and a surface brightness high enough to see structure in the source clearly. It is important to note that this selection was done before the calculation of the flood-fill sizes reported in section 7 of \cite{hardcastle23} and thus before the completion of the final value-added catalogue; it is purely based on the sizes of the components detected using PyBDSF. As a result, some of our targets have a catalogued largest angular size of $<90''$ in the \cite{hardcastle23} catalogue. In total, after rejecting some sources that were not correctly assigned a large size or an optical ID, this gave us 9,985 visually inspected objects.

We later selected only those sources with a `good' redshift (as described by \citealt{hardcastle23}) which gave a sub-sample of $7,613$ sources. Redshifts come mainly from the SDSS Data release 16 \citep{blanton17} for spectroscopic redshifts, with a small addition from the DESI early data release \citep{juneau22} and the HETDEX survey \citep{mentuchcooper23} and are otherwise estimated using the methods described by \cite{duncan21}. Radio luminosities and physical sizes are then calculated from the redshifts, the radio flux and the total angular size for each source assuming $H_0 = 70$ km s$^{-1}$ Mpc$^{-1}$, $\Omega_M = 0.3$ and $\Omega_\Lambda = 0.7$.  

We developed a morphological classification system composed of three parts: (1) an initial classification based on a `best effort' interpretation of the morphological type (such as FRI, FRII, narrow-angle tail etc), (2) a list of morphological features present (such as wings, multiple hotspots), and (3) some statement of the surrounding environment where possible (such as location within a cluster). The features we used and the source counts are summarised in Table~\ref{tab:sources} along with a `derived catalogue' section which can be used to identify subclasses of objects based on combinations of applied classifications.

These labels were qualitative and a brief description is as follows: distinct classes were FRI (Fanaroff-Riley Type I: \citealt{fanaroff74}) sources, which are centre brightened; FRII (Fanaroff-Riley Type II) sources, which are edge brightened; Single-sided jets which could be head-tail or narrow-angle tail sources; `Hybrid' or uncertain structures which may fall into multiple categories: if not paired with any other classification these are likely extended sources with a strong point-like component, odd radio circles \cite[e.g.,][]{norris21}, or unknown populations but if paired with FRI / FRII tags form a sub-population that apparently shows features of both FRI and FRII-style morphologies; possible Spirals; and Relaxed Doubles, which may be remnant sources where no jet remains visible and whose old lobes may show evidence of losses due to adiabatic expansion. 

Morphological features of interest included curvature (whether s- or c-shaped); jet misaligned from central axis (in double-lobed radio sources); wings or extended emission (in either FRI, FRII or other sources); x-shaped sources (where a pair of wings emerges in opposite directions from the lobes, and which are almost always FRIIs); straight jets; multiple or complex hotspots; continuous visible jets; banding in the lobe; one-sided jets in FRII double lobes where only one jet is visible, possibly due to relativistic beaming; and restarted sources (`double double' radio lobes or evidence of jet reorientation, particularly in straighter sources). Environmental indicators included evidence of being in a cluster (we tended to be generous with this label, as discussed below); evidence of mergers or recent disturbance (difficult to observe); and the existence of diffuse emission (e.g., radio emission not obviously associated with any source). While diffuse emission were usually already checked for association with sources beyond the field of view, environmental indicators were subjective and constrained only to immediately obvious features such as shocks, remnants or extended emission adjacent to, or apparently connected with, the source in question. The only exception to this was in the case of giant radio galaxies \citep{oei23}, most of which did not form part of this survey.

There is inevitably a degree of subjectivity in our morphological classifications. This is a particular problem in listing the effect of environment and clustering on the morphologies.   As an example, the peculiar radio galaxy J0011+3217 \citep{shobha24} was included in our sample. The field of view of our LoTSS cutout was $\sim 1'$, and was too small to reveal that this source resides within or near a cluster. On closer inspection, \cite{shobha24} found that this radio galaxy is located approximately $9.9'$ from the centre of cluster Abell 7. Given the large sample of objects assessed within this study, we were unable to consider any information beyond that visible within our $\sim 1'$ cutouts and are therefore likely not to have recognised cluster environments in many more cases.  Additionally, groups of close sources may have been incorrectly classified as clusters since we did not take account of their redshifts. Usually, if a group of sources appeared associated with one another in projection, we decided to label them as clusters worthy of follow-up investigation. Given the comparatively small number of sources classified as clusters, this subsample should be used with caution.

Most morphological labels in list (1) were considered to be mutually exclusive (apart from rare FRI / FRII hybrid sources), whilst sources could possess any number of features from lists (2) and (3). At the same time we recognised that ours is a subjective classification only, and it is sometimes difficult to be certain based on visual inspection of images with limited resolution. Additionally, we used the category `relaxed doubles' for sources where no visible jet structure was present, whilst recognising there may be some overlap between these and poorly resolved FRIIs. For the sake of consistency, visually bright sources were assumed to be poorly resolved FRIIs rather than relaxed doubles\footnote{This choice was made since relaxed doubles are comparatively rare \citep[e.g.,][]{mahatma23} and tend to be low power, perhaps as a consequence of radiative losses and adiabatic expansion.}. It is worth pointing out that it is difficult to identify multiple hotspots in sources observed at low resolution, so if a region seemed unusually bright over an extended area we flagged it as a possibility for finding evidence for multiple hotspots in future, since higher resolution images may become available for these sources in future LOFAR projects. 

Visual inspection was carried out using static images generated from the LOFAR data in a manner similar to that used for the RGZ(L) project. A simple text-based interface was used to classify each source. It is important to note that no indication of source physical properties (such as luminosity or redshift) was visible at the time of classification, in order to avoid bias (such as making assumptions about source type due to size or luminosity, particularly when those beliefs have arisen from conclusions drawn by smaller and less sensitive surveys).

The final output of the classification process is a catalogue that is released with this paper \footnote{\url{https://lofar-surveys.org/releases.html}}. It consists of the entries for our target sources from the value-added source catalogue of \cite{hardcastle23} (version 1.1) together with boolean flags representing our various classifications. 

\begin{table*}
    \caption{List of primary categories (column 1), available options (col 2), number of sources in each category (col 3) and number of sources in category with usable redshift (col 4) and usable host galaxy mass estimator (col 5).}
    \label{tab:sources}
    \centering
    \begin{tabular}{l | l | r| r | r }
         Type & Classification & No. Sources & $z_{best}$ &Mass\\
         \hline 
         \multirow{5}{*}{Initial classification}
         & FRI & 2992 & 2406 & 1417\\
         & FRII & 6231 & 4693 & 2498\\
         & Hybrids & 966 & 751 & 388\\
         & Spirals & 234 & 145 & 77\\
         & Relaxed Doubles & 440 & 361 & 239 \\
         \hline 
         \multirow{11}{*}{Morphology}
         & C-curvature & 2379 & 1906 & 1114\\
         & S-curvature & 1082 & 864 & 491\\
         & Misalignment & 1287 & 1040 & 588\\
         & Wings & 1564 & 1275 & 722\\
         & X-shaped & 179 & 148 & 79\\
         & Straight jets & 358 & 269 & 128\\
         & Multiple hotspots & 2040 & 1613 & 916\\
         & Continuous jets & 337 & 279 & 143\\
         & Banding & 133 & 96 & 49\\
         & One-sided & 23 & 19 & 13\\
         & Restarted & 193 & 149 & 78\\
        \hline
        \multirow{6}{*}{Environment} 
        & Cluster & 72 & 54 & 28\\
        & Merger & 63 & 49 & 25 \\
        & Diffuse emission & 111 & 75 & 43\\
        & Unknown & 126 & 100 & 46\\
        \hline 
        \multirow{6}{*}{Derived Catalogue} 
         & Compact sources and other hybrids & 373 & 278 & 134 \\
         & Hybrid FRI/FRII & 593 & 473 & 254 \\ 
         & Curved FRIs & 1791 & 1447 & 858\\
         & Curved FRIIs & 917 & 730 & 413\\
        & Straight \& multi hotspots & 121 & 98 & 52 \\
         \hline 
         \end{tabular}
    
\end{table*}

\section{Initial classifications}

In this section we discuss the results of the classification process, for convenience combining classifications that will be discussed together in later sections of the paper. Not all of the classifications are discussed here as not all are relevant to, or included in, the later parts of the paper.

\subsection{Fanaroff-Riley classifications}
\begin{figure}
\includegraphics[width=1\linewidth]{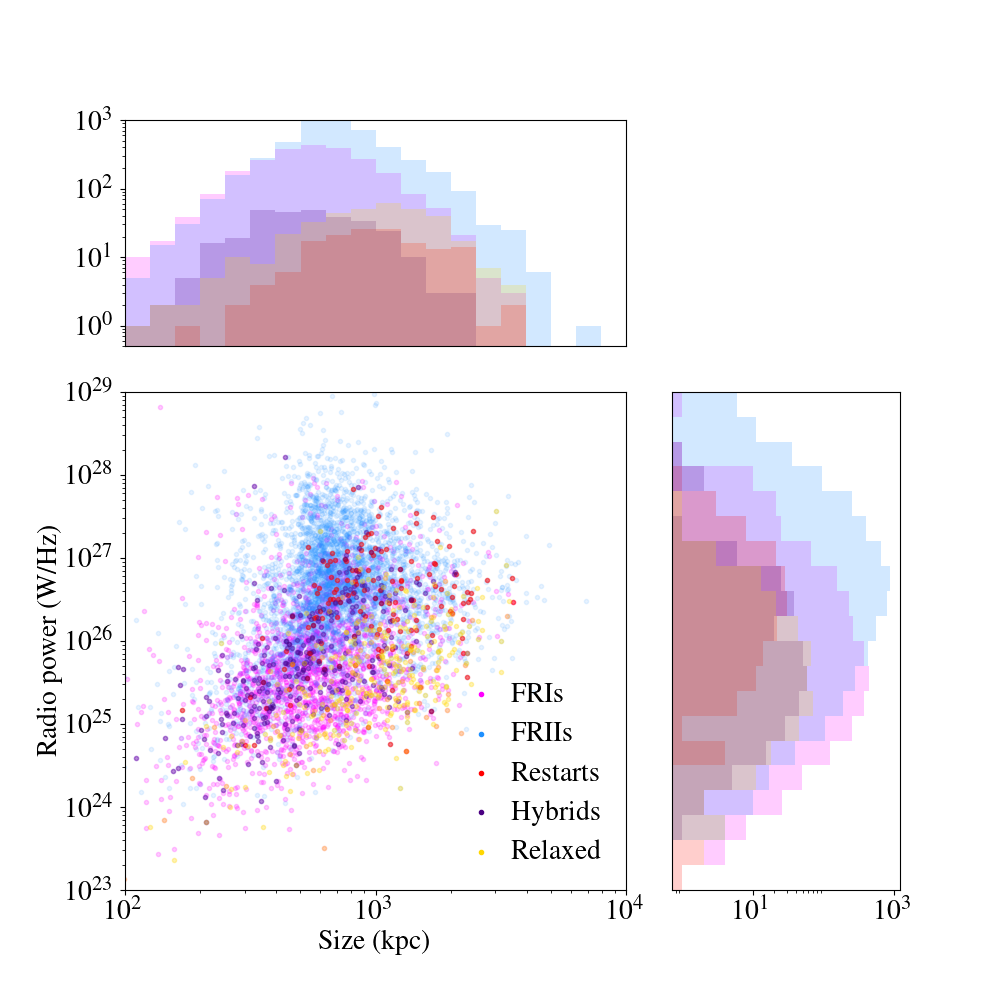}
\caption{Power-linear size ($PD$) diagram showing the distribution of main classification types used in this paper. These include FRIs and FRIIs (magenta and blue), morphologically ambiguous `hybrids' (purple), restarts (red), and relaxed doubles (yellow). The latter two represent distinct stages of the AGN lifecycle.}
\label{fig:scatter_fris_friis.png}
\end{figure}

7099 usable sources were tagged as Fanaroff-Riley Type I or II (hereafter referred to as FRIs or FRIIs respectively), meaning most of all large, powerful galaxies fell into one or both of these groups. Of these, 4693 were FRIIs while 2406 were FRIs. 751 sources could not be adequately classified based on visual inspection alone or had features associated with both galaxy types. These will be discussed in the following section; however, they have not been removed from this analysis. Given the large angular size and flux cut used for this classification, no sources were considered likely to be FR0 and this classification has not been included. Throughout the paper, we use the power-linear size diagram ($PD$ diagram) to visualise the radio properties of subsamples of objects, and Fig.~\ref{fig:scatter_fris_friis.png} shows the distribution of FRIs and FRIIs in this dataset, with a central population of `hybrid' sources emerging (see following section). As usual, we see evidence for a FRI-FRII `break' at 144-MHz radio luminosities around $10^{26}$ W Hz$^{-1}$, but as previously reported \citep{mingo19} it is not at all sharp, with many objects classified as FRIIs lying below this luminosity and a certain number of FRIs lying above it.

\subsection{Hybrid FRI / FRIIs}
\label{sec:hybrid}
\begin{figure*}
\includegraphics[width=1\linewidth]{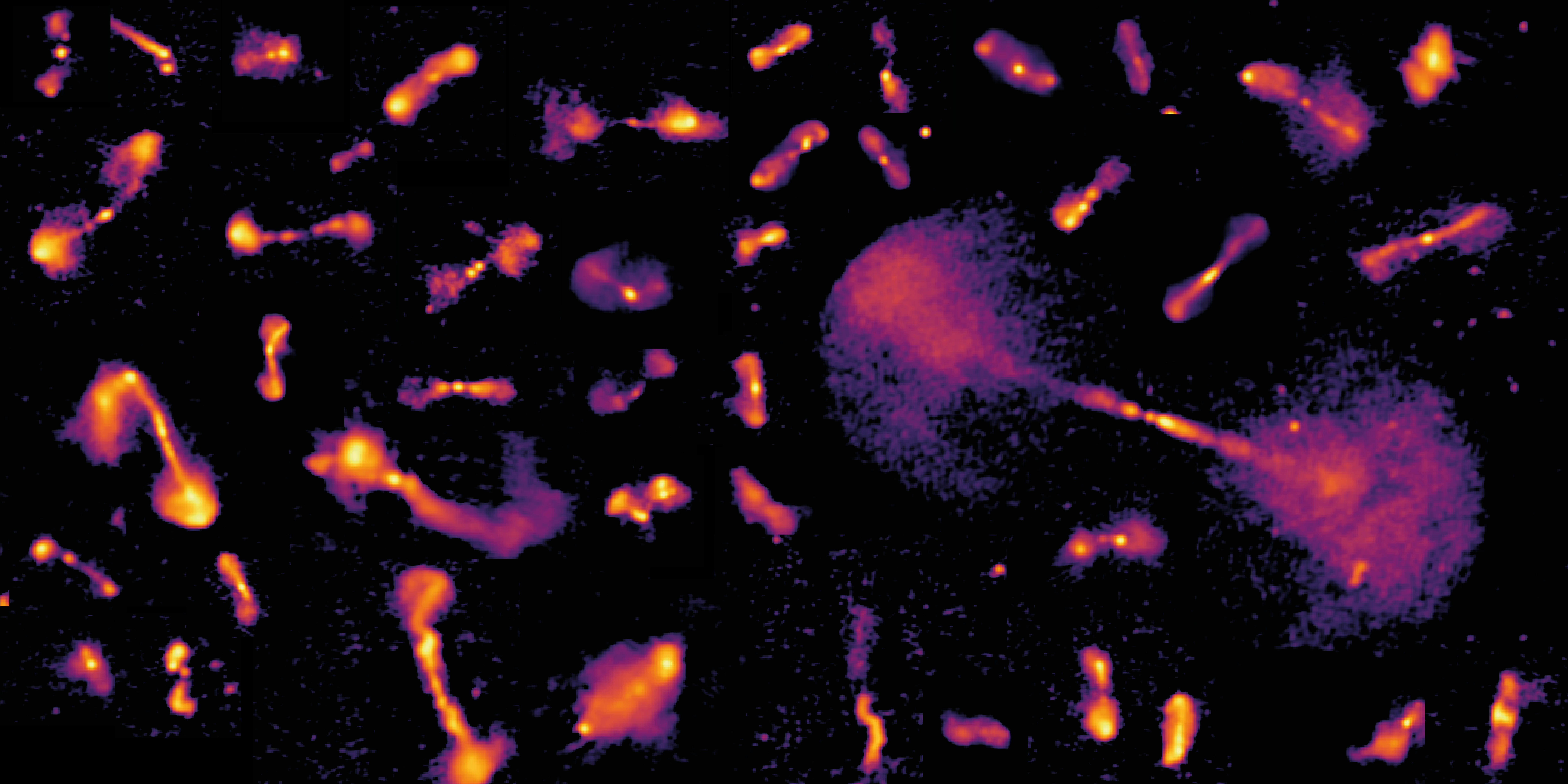}
\caption{Montage of representative sources tagged as appearing to have features of both FRIs and FRIIs (based on visual inspection only). Images are from LoTSS DR2 and are all on the same angular scale at $6''$ resolution.}
\label{fig:hybrids.png}
\end{figure*}

This category emerged from the classification due to a small number of sources ($<0.5\%$) that had features of both FRI and FRII jet morphologies: for example, bow-shaped brightened centres with terminal hotspots, or classic double-lobed galaxies with a bright central core, or some other unusual feature (see discussion for details). These galaxies were tagged with both FRI and FRII labels, alongside the `hybrid / other' option to create a distinct subclass. Some examples of this category are shown in Fig.~\ref{fig:hybrids.png}.

Figure~\ref{fig:scatter_fris_friis.png} shows the distribution of hybrid sources with respect to jet power and angular size (hybrid sources appear as dark purple). While they can appear anywhere, it is interesting that most fall between the main FRI and FRII subpopulations. They may thus form a distinct category of sources that represent a stage of evolution between one jet type and another. Some of these objects may be so-called wide-angle tailed radio galaxies, which exhibit characteristics of both FRI and FRII \citep{hardcastle04} while others may be transitional objects in other ways \citep{mingo19}.

The nature of hybrid FRI/FRII sources, or HyMoRS \citep{Gopal-Krishna+Wiita00}, has been under discussion for some time. Recently, \cite{mingo19} discussed the possibility of some jets in LOFAR-selected samples having centre-brightened, FRI-type morphology on one side and edge-brightened, FRII-type morphology on the other side, resulting in heteromorphous emission. While the possible mechanisms behind such possibilities remain unclear, many such hybrid sources are present in the sample that we have inspected.

However, we are using the hybrid term more broadly. In our classification, the bulk of large jets in the hybrid class were those that exhibited both edge and centre-brightened features symmetrically, particularly in situations where strong shocked lobes and terminal hotspots appeared to be present (e.g., many of those observed in Fig.~\ref{fig:hybrids.png}). We did not exclude asymmetrical objects, so objects that would be classed as HyMoRS are included alongside sources that have FRI-like cores and FRII-like lobes.

Our hand-selected sample of hybrids often falls between FRIs and FRIIs in terms of size, luminosity and host galaxy mass (see Fig.~ \ref{fig:scatter_fris_friis.png} and the discussion in Section \ref{sec:mass}). This suggests that so-called hybrids could indeed be a separate population with distinct characteristics. More work is required to explore the environments of this subpopulation and, therefore, whether or not they are more likely to be suitable host candidates for true supermassive binary black hole systems. It would also be interesting to see whether asymmetrical HyMoRS sources exhibit the same properties as the more general hybrids shown here.

\subsection{Spirals, clusters and diffuse emission}
\begin{figure*}
\includegraphics[width=1\linewidth]{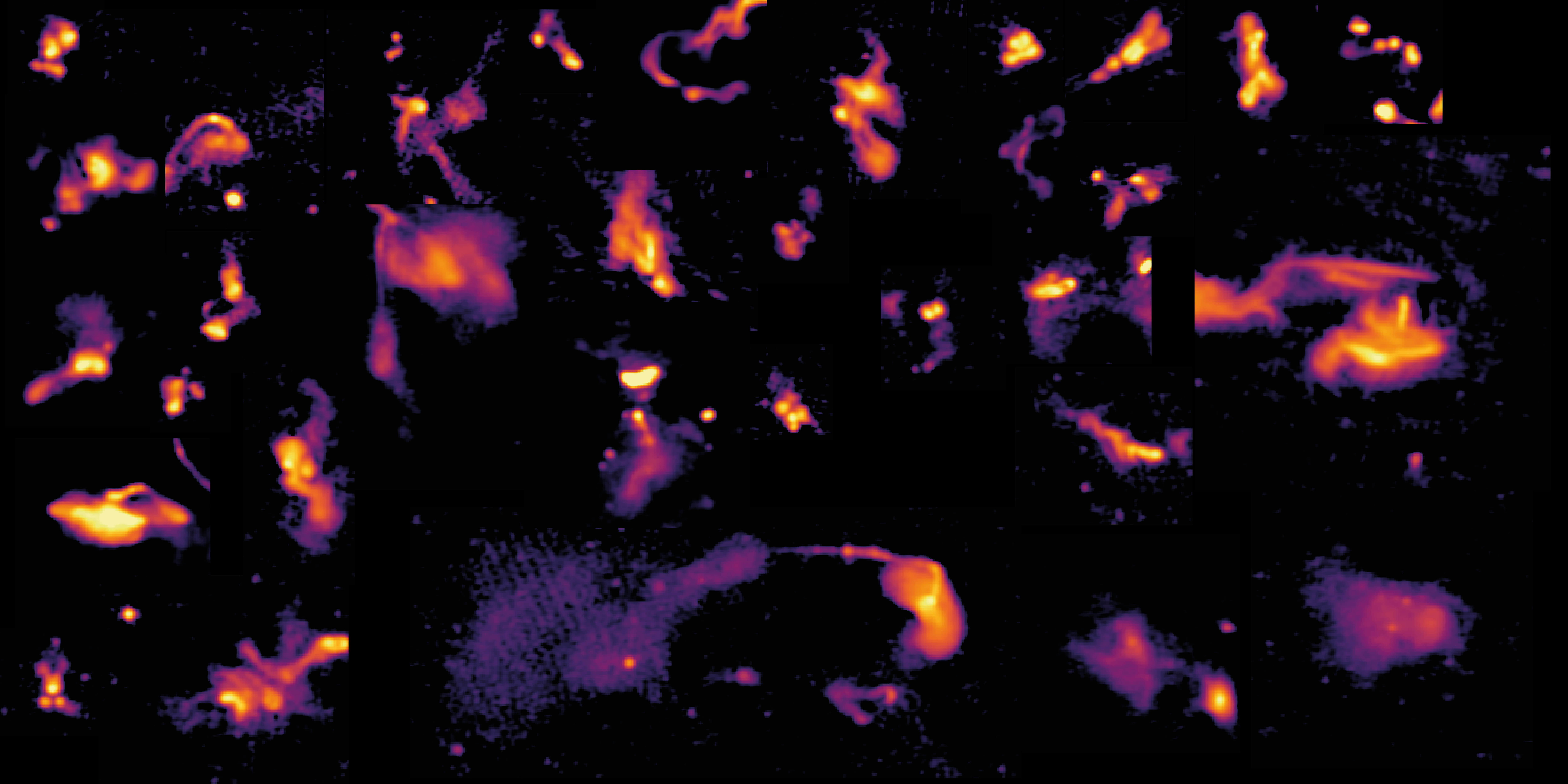}
\caption{Montage of representative sources tagged as `clusters'. Details as in Fig.\ \ref{fig:hybrids.png}.}
\label{fig:clusters.png}
\end{figure*}
Clusters have a long association with complex emission. \cite{vanweeren19} show many sources that contain diffuse, relic and extended emission that are not always directly linked to a specific cluster, but are caused by merger histories, shocks and fossilised plasma. These include `radio phoenixes' where old emission is reactivated by a merger or shock, such as Abell 1033 \citep{degasperin15}, which is included in Fig.~\ref{fig:clusters.png}. The complex nature of cluster emission and associated processes poses a challenge for detecting supermassive binaries through precession morphology due to unpredictable effects on emission, despite clusters arguably being a good place to search for binaries.

Many of the objects tagged as spirals and clusters have poor quality images. Due to this, further analysis may reveal that these were not accurate classifications. Table~\ref{tab:sources} shows that 234 sources have been labelled as possible spirals, where the visual selection criteria were (1) emission that showed features that were clearly associated with spirals or (2) more blurred emission that is consistent with being a spiral galaxy, such as an elongated oval axial ratio. Of these, 145 had plausible optical IDs. 

72 sources were labelled as having characteristics of clusters (where the selection criteria were (1) multiple radio sources close together and (2) evidence of highly disturbed jets or possible interactions between sources). This was done in the context of jet precession rather than being a direct search for clusters in their own right, and so our selection of clusters is limited to the regions immediately adjacent to the source and thus neither uniform nor complete. On further investigation, objects tagged as clusters (some examples of which are shown in Fig.\ \ref{fig:clusters.png} usually showed distinctive cluster environments (for example, known Abell clusters or groups where recognisable from radio emission). Unfortunately, given the limited field of view of the classifications, it is likely that less obvious clusters existing more than a few arcminutes from the target position would not have been identified through this method; many other sources in the sample may exist in clusters or be adjacent to them. 

\subsection{Restarted sources}
\begin{figure*}
\includegraphics[width=1\linewidth]{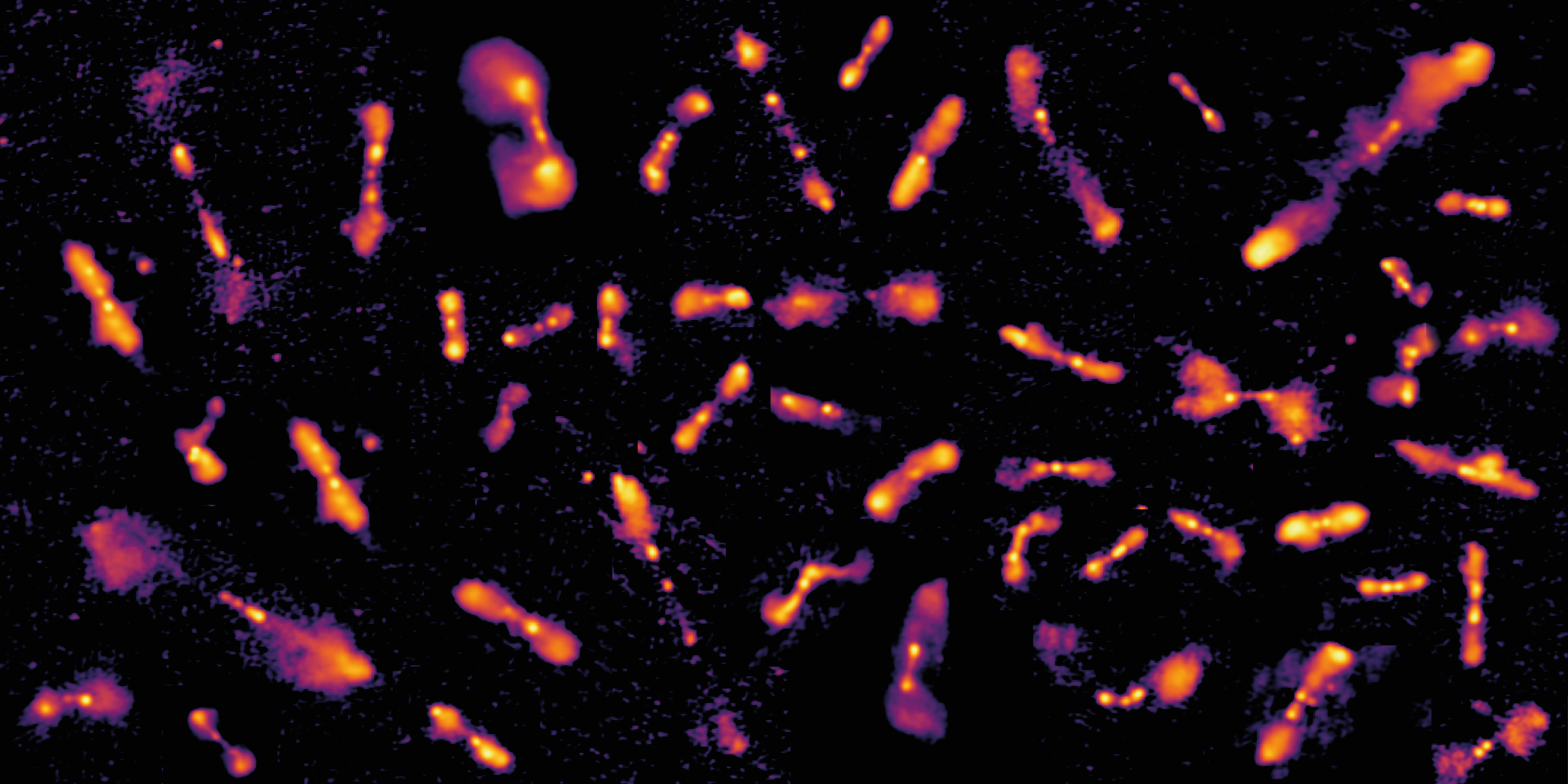}
\caption{Montage of representative sources tagged as `restarts'. Details as in Fig.\ \ref{fig:hybrids.png}.}
\label{fig:restarts.png}
\end{figure*}

\begin{figure}
\includegraphics[width=1\linewidth]{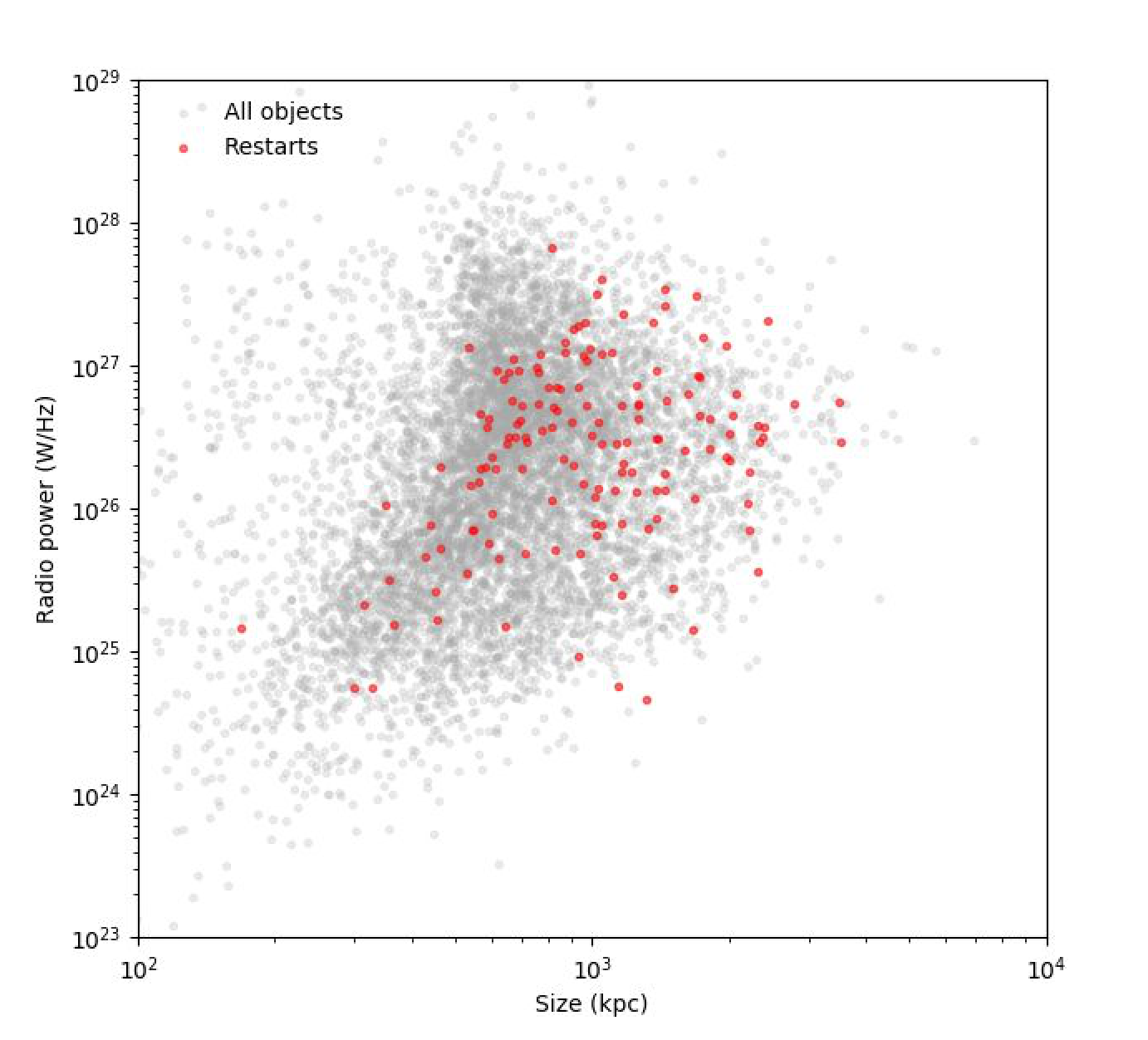}
\caption{$PD$ diagram showing population of sources tagged as restarted sources. Red sources represent restarted sources while grey sources represent the full population.}
\label{fig:scatter_restarts.png}
\end{figure}

Fig.~\ref{fig:restarts.png} shows a collection of sources labelled as 'restarted' sources. The morphological criteria for this category were any FRII-like jet (or, rarely, FRI) showing evidence of (1) jets that appear shorter and brighter than the surrounding lobe material (2) terminal hotspots in the centre of lobes, particularly when connected by strong jets, or (3) apparent reorientation of the jet in a way that is distinct from, or is in combination with, precession signatures. As with all categories, these criteria are subjective and non-exclusive. The category includes, but is not limited to, classical double-double radio galaxies \citep[e.g.,][]{schoenmakers00b,mahatma19}.

\subsection{Compact sources and other hybrids}
\label{subsec:pointsources}
\begin{figure}
\includegraphics[width=1\linewidth]{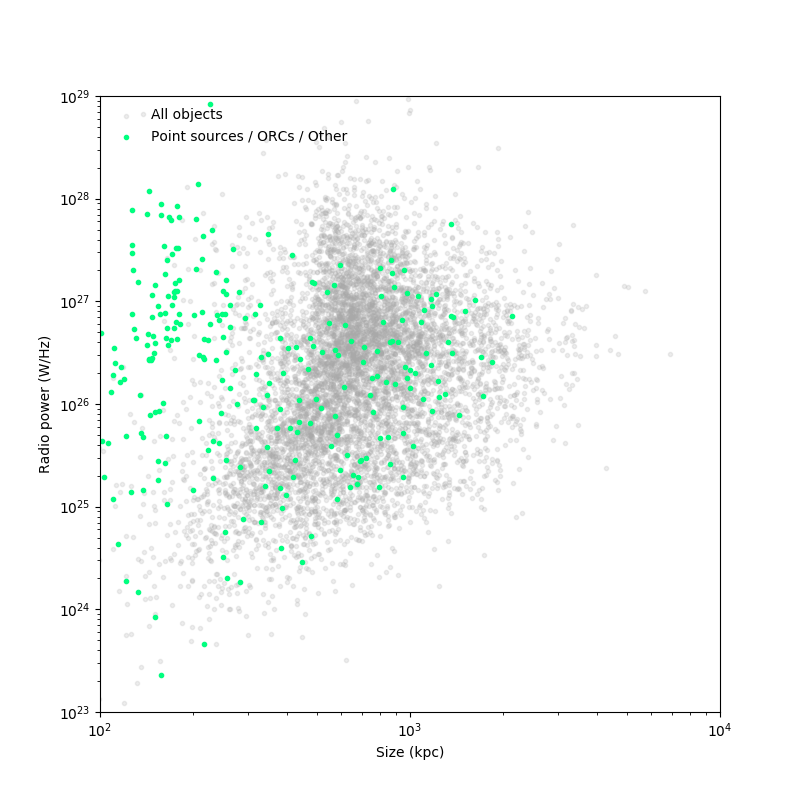}
\caption{$PD$ diagram showing population of sources tagged with `hybrid' features that also do not show any evidence of jets. Green sources represent these objects while grey sources represent the full population.}
\label{fig:scatter_pointsources.png}
\end{figure}
\begin{figure*}
\includegraphics[width=1\linewidth]{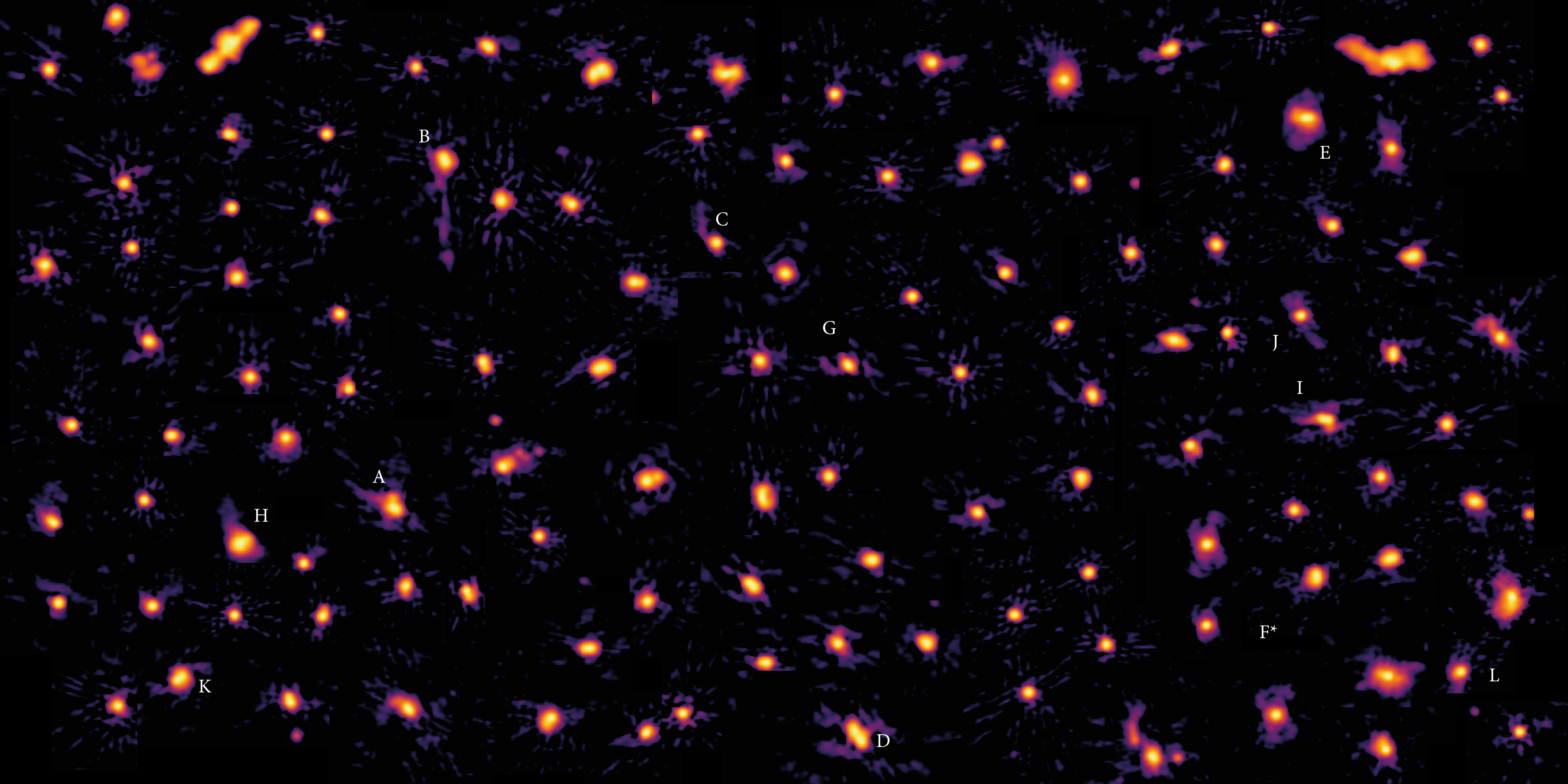}
\caption{Montage of representative sources tagged as `other' or hybrid sources that did not otherwise have associated jets. This population is filtered to only include small, bright sources above sizes of $10^{26}$ W Hz$^{-1}$ and with sizes below 500 kpc. Images are all on the same angular scale.}
\label{fig:pointsources.png}
\end{figure*}

The hybrid / other category was used for multiple purposes. As previously mentioned, when associated with both the FRI / FRII category, it was used to denote resolved jets with complex structure. However, when selected alone, it was used to represent emission dominated by point sources with no other classification. This was necessary because a population of very bright point sources surrounded by apparent extended emission only became obvious once the classification had begun. Fig.~\ref{fig:pointsources.png} shows a representative selection of these objects while Fig.~\ref{fig:scatter_pointsources.png} highlights the unusual distribution of these objects. Many of these are likely to be unresolved double-lobed galaxies or spirals whose morphology may become apparent with higher resolution image maps. 

Of particular interest is the subset of 119 small, high luminosity sources, which have sizes in the \cite{hardcastle23} catalogue below 500 kpc and have a radio power of more than 10$^{26}$ W Hz$^{-1}$. Fig~\ref{fig:scatter_pointsources.png} shows this population sitting outside of the main distribution of our targets on a $PD$ diagram. Representative images from this subset are shown in Fig.~\ref{fig:pointsources.png}. 

Many of these sources appear to have unusually high luminosity, given their size. This population deserves further study due to the possibility of unusual and complex emission associated with some of these sources. Unfortunately, due to their limits in angular resolution, it is sometimes difficult to tell what is real emission and what are imaging artefacts (particularly when unresolved and smaller than the beam size). For example, sources A, B and C in Fig.~\ref{fig:pointsources.png} appear to show point-like sources or compact double-lobed structures with plumes or tails which look like genuine structures, whereas several sources around F* show near-identical artefacts most likely due to ionospheric errors in the PSF. This is made even more likely since those particular subset of sources appear sequentially and were observed in the same field, and in some cases the same facet. Their small physical size, relative to other objects in the sample, is likely the result of a disagreement between PyBDSF and the flood-fill algorithm about the reality of the extended emission that surrounds them.

Naturally, most objects in this population are ambiguous and follow-up observations are likely to be required to separate what is true emission from what might be due to artefacts.

Some sources in this population may be `odd radio circles', \cite[e.g.,][]{norris21}, but we have not specifically looked for this category. Very few clear examples of this type of object appear in our sample. They have generally been identified in ASKAP observations with significantly lower resolution than the $6''$ observations that we use here, and so it is possible either that they are resolved out in our observations or that they exhibit more structured morphology at higher resolution.

It is worth noting that `FR0' galaxies \citep{baldi16,garofalo19}, which are core-brightened sources similar to FRIs without visible extended structure, are largely excluded from our sample due to our size cutoff in selection. However, our population of bright compact objects would be worth revisiting within that context.

\subsection{Relaxed doubles}
\label{sec:relaxed}
\begin{figure*}
\includegraphics[width=1\linewidth]{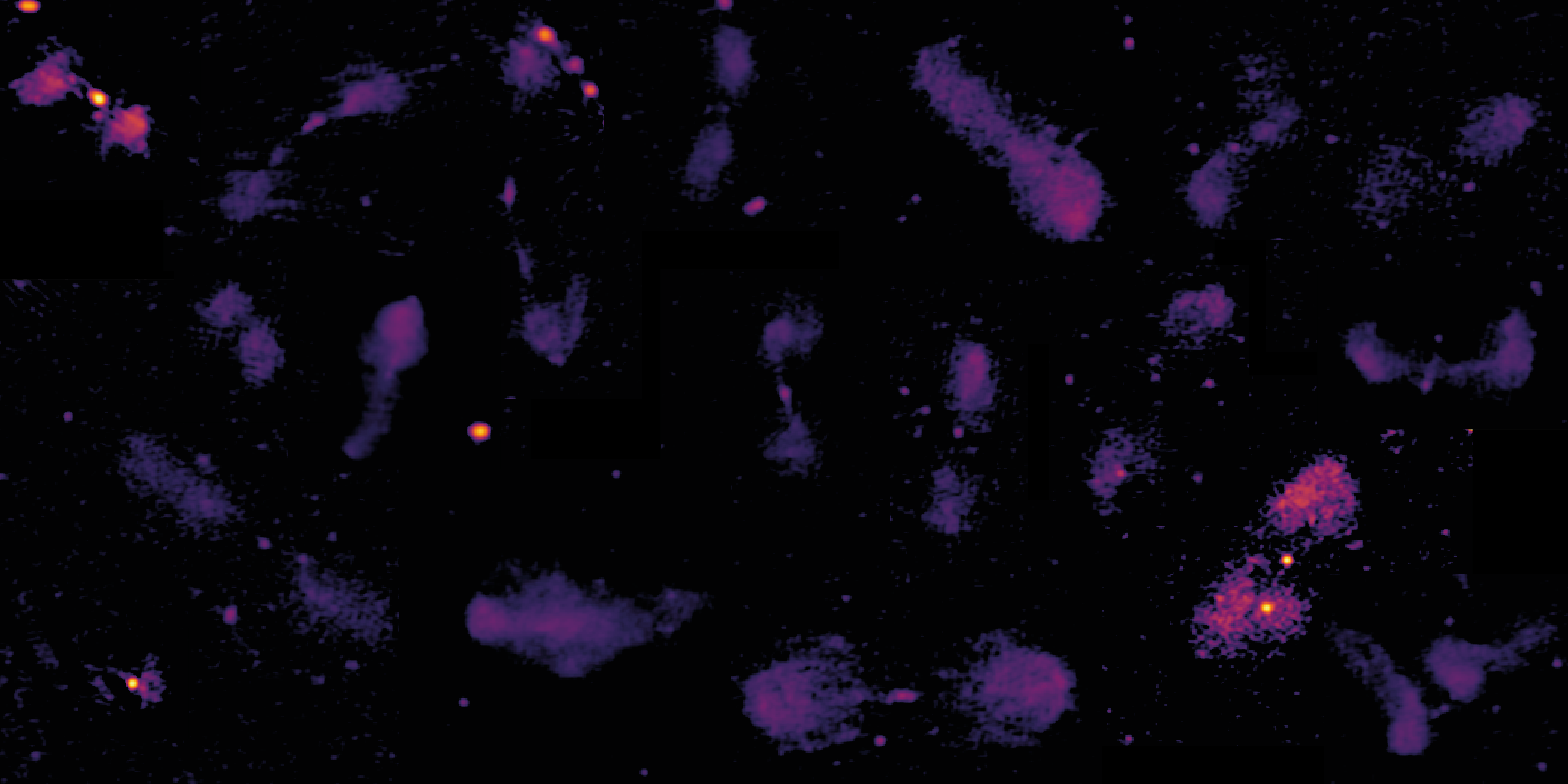}
\caption{Montage of representative sources tagged as relaxed doubles. Details as in Fig.\ \ref{fig:hybrids.png}.}
\label{fig:relaxed.png}
\end{figure*}

The selection criterion for relaxed doubles is any source (typically, but not exclusively, an FRII) which shows few or no features and appears to be exhibiting signs of radiative losses and adiabatic decay (low surface brightness and expanded lobe structure, with no visible jets or hotspots). Fig~\ref{fig:relaxed.png} shows a sample of these sources. Many of these would have been classified in previous work as remnant radio galaxies \citep{Parma+07,Brienza+17}.

\begin{figure}
\includegraphics[width=1\linewidth]{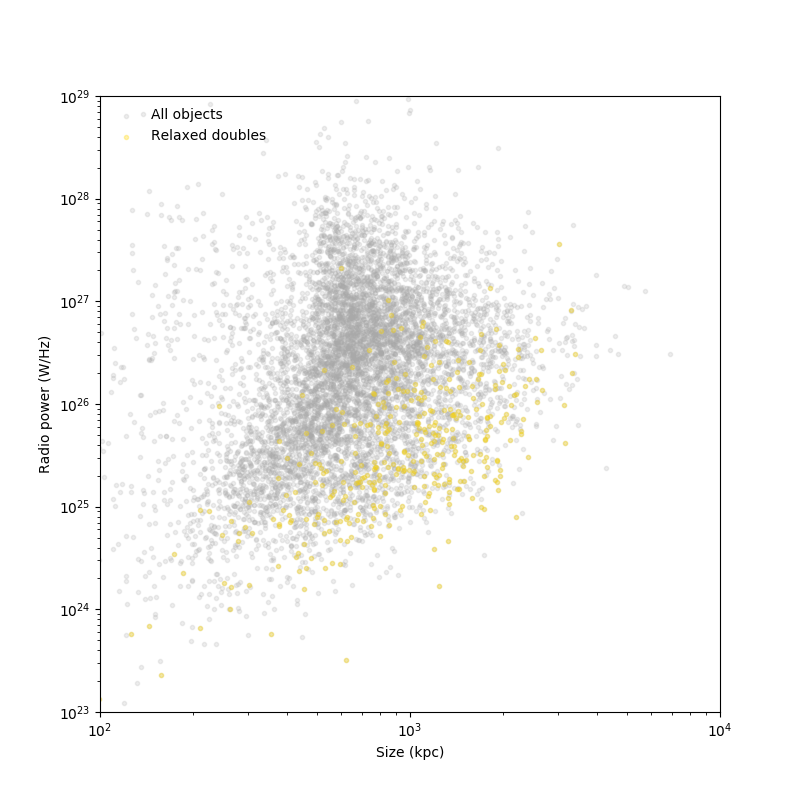}
\caption{$PD$ diagram showing population of sources tagged with `relaxed' features. Yellow sources represent relaxed doubles sources while grey sources represent the full population.}
\label{fig:scatter_relaxed.png}
\end{figure}

\subsection{X-shaped sources}\label{subsec:xshaped}

\begin{figure*}
\includegraphics[width=1\linewidth]{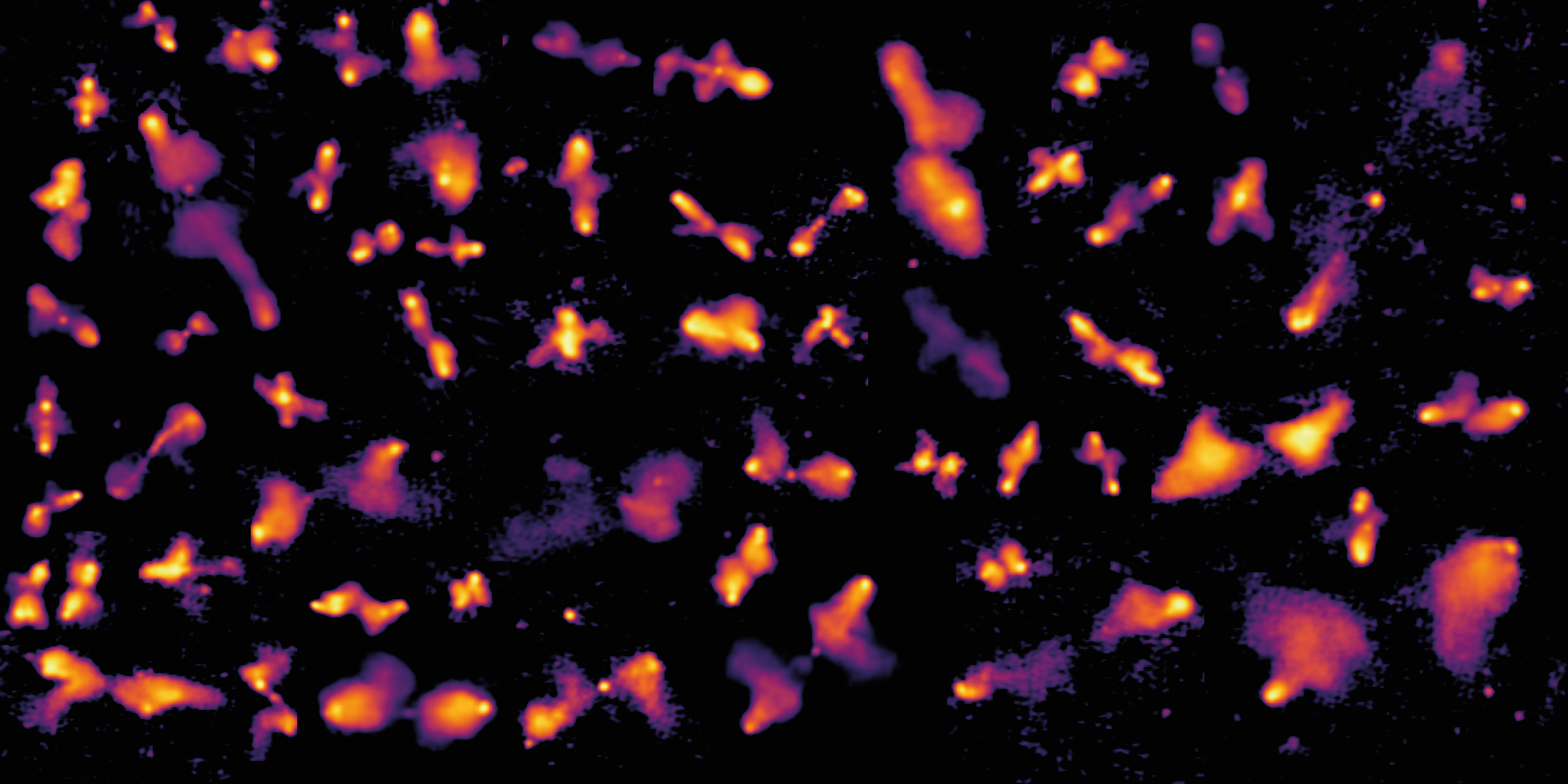}
\caption{Montage of representative sources tagged as `x-shaped'. Details as in Fig.\ \ref{fig:hybrids.png}.}
\label{fig:xshaped.png}
\end{figure*}

X-shaped sources are interesting because their low surface brightness wings can be used as an indicator for precession \citep{ekers78}, realignment, and potentially restarts (depending on the flow dynamics of the source involved). While rapid realignment is a long-standing explanation for this type of object \citep{merritt02}, \cite{horton20b} found that x-shaped sources could be generated by precession due to projection effects along certain lines of sight. 96 x-shaped sources (50\%) had at least one precession indicator from the indicators discussed in Section \ref{sec:precession}; 19 (10\%) had all three. This is a substantially higher fraction of precession indicators than the general population.

Other explanations for x-shaped morphology exist, however. They may simply be associated with backflow in asymmetrical environments \citep[e.g.,][]{gourab22, cotton20,dennetthorpe02}. The backflow model has long been proposed and examined extensively in the literature as an alternative to jet reorientation \citep{leahy84,dennetthorpe02}. In such a case deeper LOFAR observations may reveal larger-scale structures that are inconsistent with a pure precession or realignment model \citep{hardcastle19}. Alternatively, since there is considerable evidence that massive galaxies at high redshift are products of merging between lower-mass galaxies such sources at high redshift \citep[e.g.][]{barthel88}, some X- or cross-shaped morphologies could be indicators of radio sources that are produced by two galaxies in the process of merging, each of which have radio-loud jets associated with different SMBH axes. In such a scenario, the statistics of future high resolution LOFAR surveys of the relative number of X-shaped radio sources as a function of redshift could provide fundamental information about the importance of such processes in massive galaxy evolution.

X-shaped sources are rare in our sample (only $\sim 2$\% of the total) but our inspection still selects a large number of objects relative to previously known samples \citep[e.g.][]{cheung07}.

\subsection{Wings and plumes}
Wings and plumes were general labels used for any extended emission from any source that extended beyond the lobe boundary. As this is a highly subjective term, many sources were included: 1564 sources (16\% of the total) were defined as having wings or plumes and 1275 of these had usable redshifts. These sources spanned a broad range of morphological types including FRIs, FRIIs, hybrid sources and others. Most relaxed doubles did not have wings or plumes; out of the 440 total identified objects classed as relaxed doubles, less than 5\% (21) were identified as having wings. This may be a surface brightness selection effect or a real feature of the morphological class.

In earlier work wings have often been interpreted as indicators of jet precession \citep[e.g.][]{gower82,hunstead84,krause18,misra23}. We find that objects that we class as winged have a modestly increased fraction of the precession indicators discussed in Section \ref{sec:precession}. 40\% of winged sources have at least one precession indicator, compared to 26\% of the non-winged population; 8\% have all three, compared to 4\% in the non-winged population. As with x-shaped morphology, it is therefore possible that the physical causes of the low-surface brightness lobe extensions are related to the physical factors that drive our precession indicators.

\section{Host galaxy mass}
\label{sec:mass}
\begin{figure}
\includegraphics[width=\linewidth]{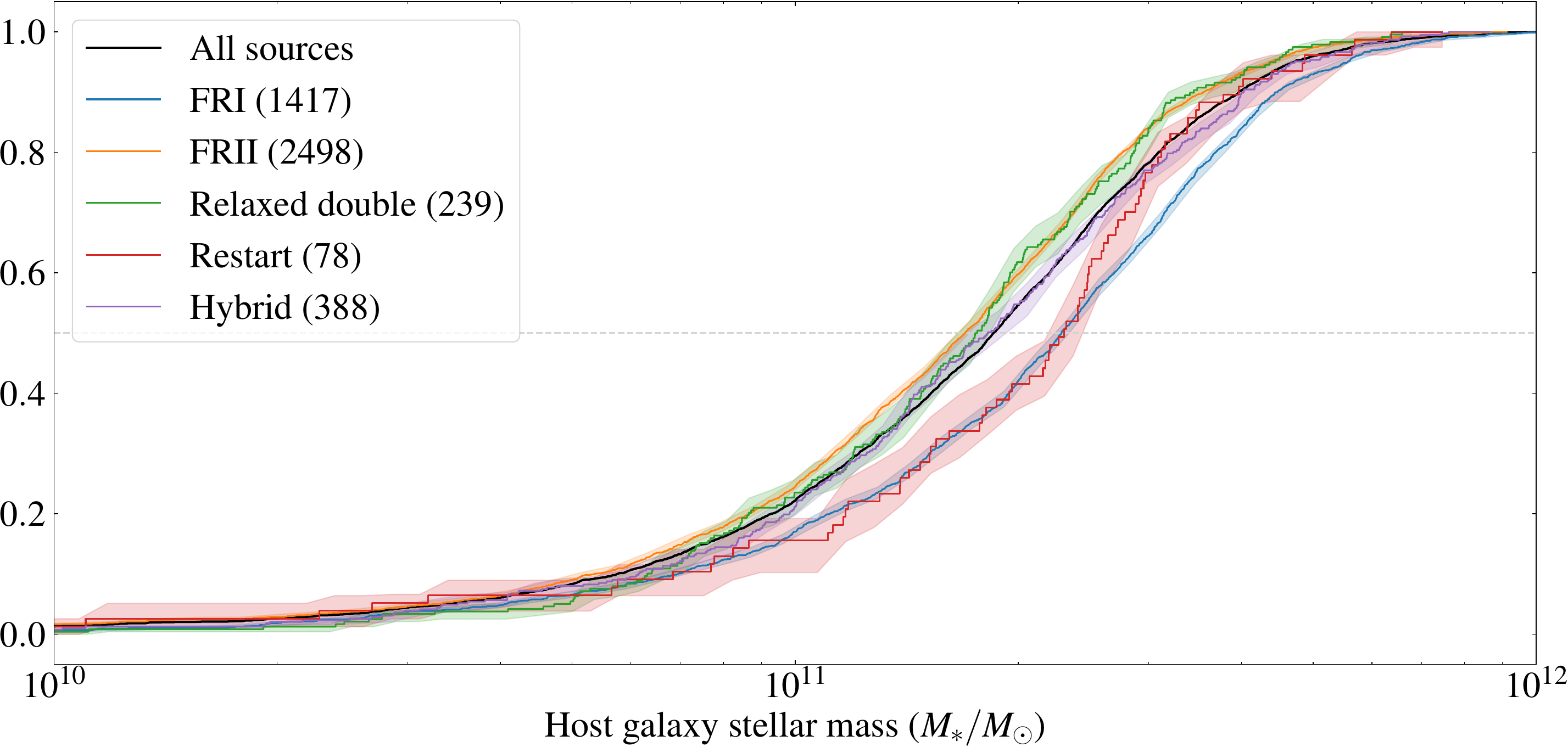}
\caption{Cumulative distribution of mass as a function of morphological indicator. The solid line shows the mass distribution for all sources in the sample and the coloured lines indicate morphological subsamples. The shaded areas show $1\sigma$ confidence intervals from bootstrap and the faint dashed line shows the location of the median.}
\label{fig:massdist-mainclass}
\end{figure}

The parent galaxy sample \citep{hardcastle23} provides host galaxy mass estimates for a large fraction of the radio sources it contains. To have a good mass estimate an object must have a reliable redshift and accurate photometry over a number of bands, and must be well fitted to a galaxy spectral template (which means that objects with a quasar host galaxy are not included). In total, 4,220 objects in our sample (out of 7,613 with a good redshift estimate) had a measured mass. These objects are likely biased to lower redshifts with respect to the sample as a whole. Nevertheless, they give us some information about the host galaxy properties for our various classes. Fig.\ \ref{fig:massdist-mainclass} shows the overall distribution of host galaxy masses and the distribution for the main morphological classifications for this sample. This plot has several interesting features. Firstly, we see that morphologically selected FRIs have host galaxies that are significantly more massive than those of FRIIs. This is consistent with much earlier work on the magnitude distributions of the two morphologically selected classes \citep{lilly84,owen+laing89,zirbel96} but shows up particularly well in our large, homogeneously selected sample. This is not a redshift effect, for although the median redshift of FRIIs is higher than that of FRIs in the overall sample, the difference persists if the median redshifts of the samples are made to match by cutting out the high-redshift FRIIs from the sample. Secondly, we see that the mass distribution of hybrid FRI/FRII sources is intermediate between that of FRIs and FRIIs, more or less as we would expect, supporting the discussion of section \ref{sec:hybrid}. Thirdly, the mass distribution of relaxed doubles closely matches that of the FRIIs, supporting the idea that the relaxed sources are largely remnant FRIIs (Section \ref{sec:relaxed}); presumably remnant FRIs rapidly fade below the surface brightness detection threshold of LOFAR. And finally, although with lower confidence due to the small sample, the restarted sources seem to have significantly higher masses than the FRII population as a whole, despite being mostly FRII-like in morphology. Whether this indicates contamination of the restart sample with e.g. wide-angle tail radio galaxies, which can mimic restarted lobes, or whether it points to a real physical difference in the hosts and/or environments of restarts, remains unclear, but this effect was not detected in earlier, smaller samples focusing on the double-double radio galaxy population only \citep{mahatma19}.

\section{Precession samples}
\subsection{Curvature in FRIs and FRIIs}
\begin{figure*}
\includegraphics[width=1\linewidth]{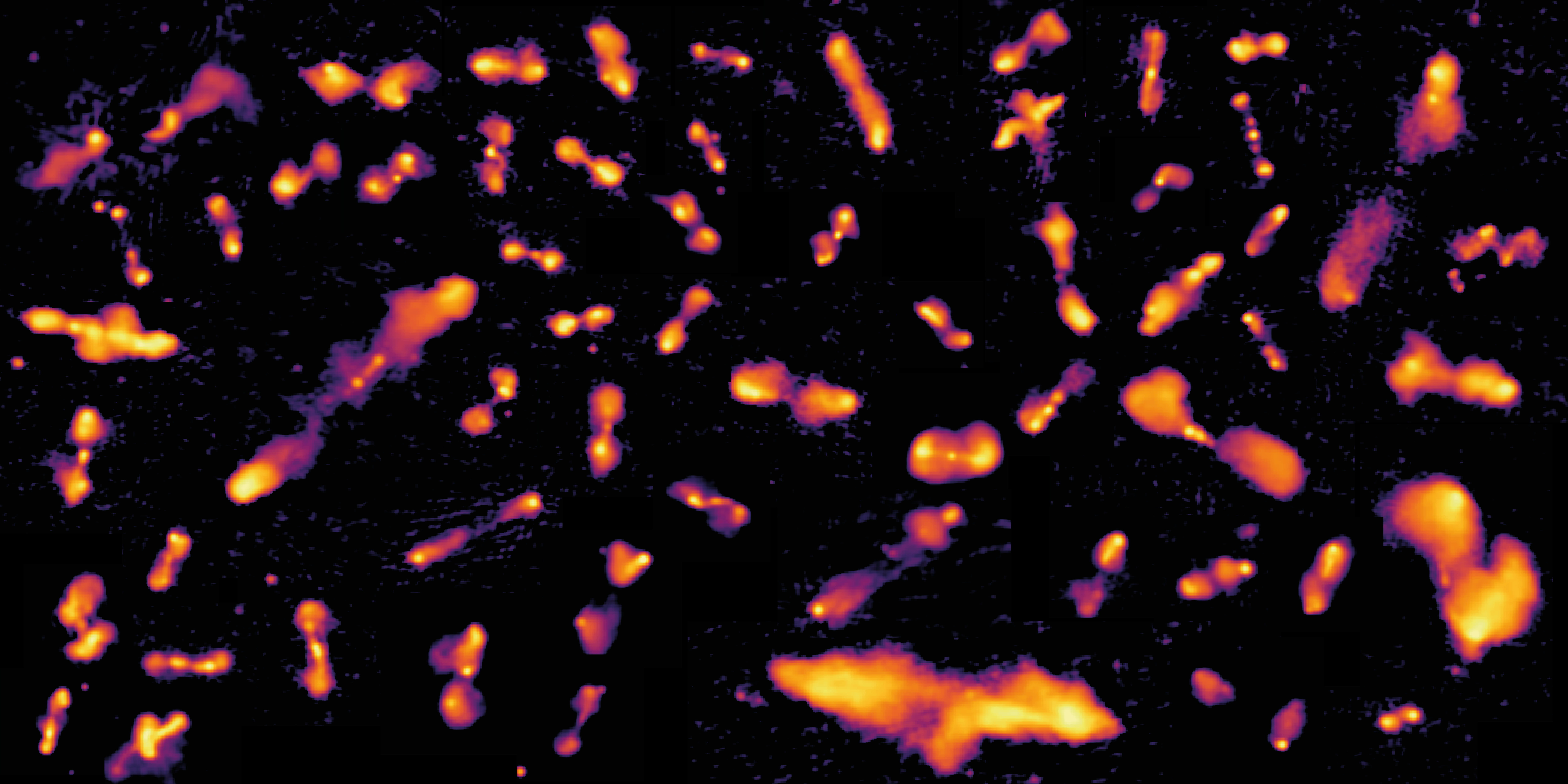}
\caption{Montage of representative sources tagged as FRIIs with characteristic s-shaped curvature. Details as in Fig.\ \ref{fig:hybrids.png}.}
\label{fig:curved-friis.png}
\end{figure*}

Jet curvature is a primary indicator of precession, particularly in the case of s-shaped curvature and FRII jets misaligned from their central axis (see Fig~\ref{fig:curved-friis.png} for examples). However, curvature can also be produced without precession, particularly in the case of FRI jets, where c-shaped curvature is seen in most sources at some level. In addition, the orientation of the jet on the plane of the sky can result in the jet appearing curved in one way or another (or straight when it might be curved; see \cite{horton20b}. 

\begin{figure*}
\includegraphics[width=1\linewidth]{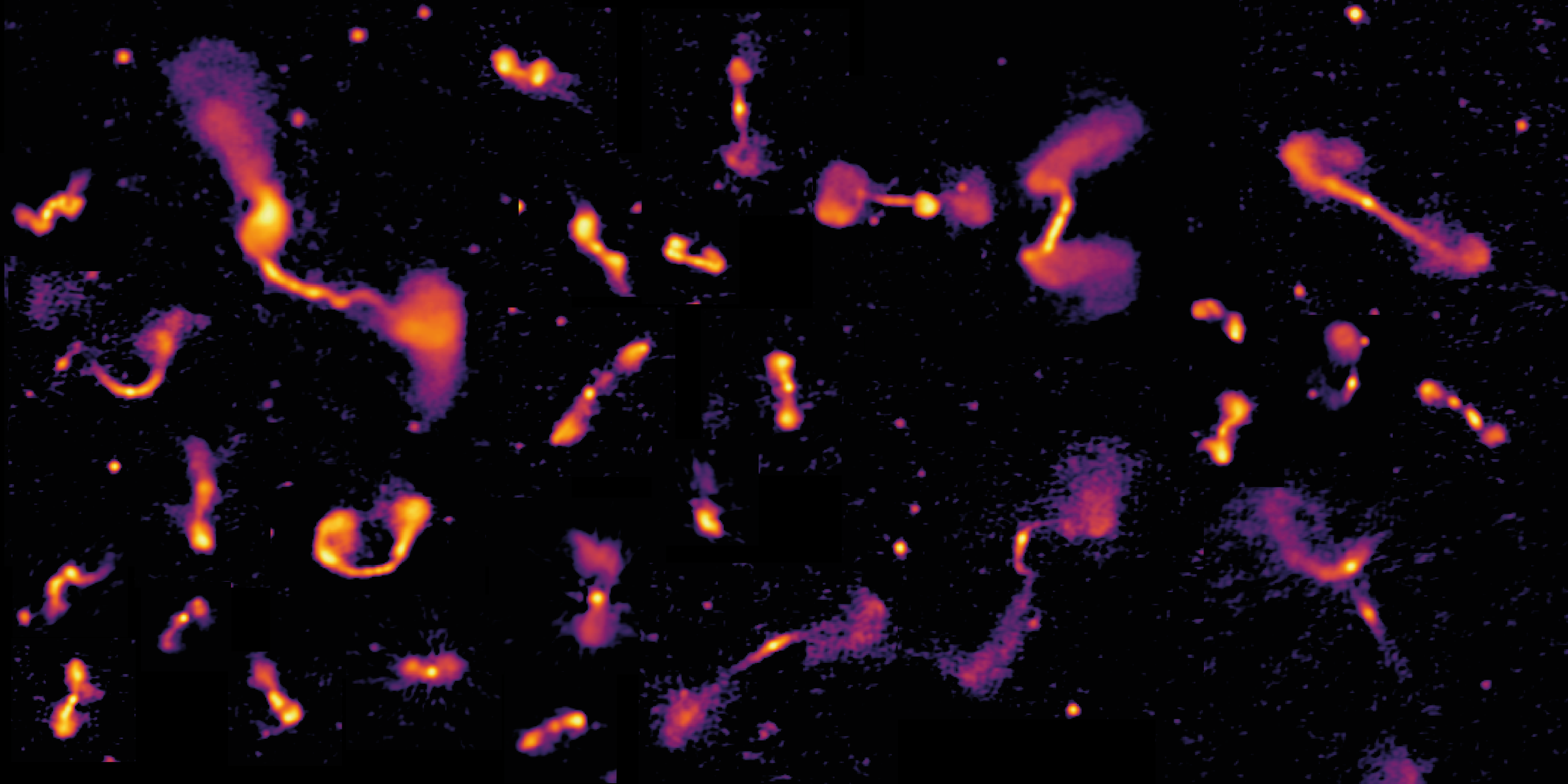}
\caption{Montage of representative sources tagged as FRIs with characteristic c-shaped curvature. Details as in Fig.\ \ref{fig:hybrids.png}.}
\label{fig:curved-fris.png}
\end{figure*}

\begin{figure}
\includegraphics[width=1\linewidth]{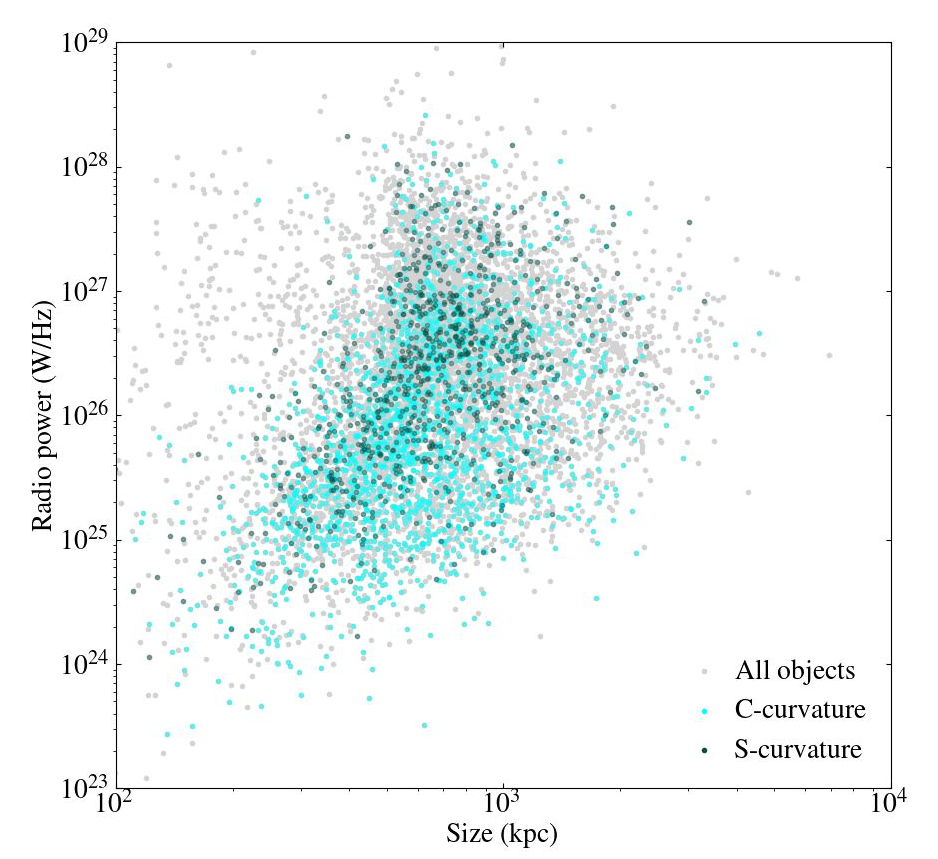}
\caption{$PD$ diagram showing population of sources with s- and c-shaped curvature. Grey sources represent the full population while sources with c-curvature are presented in light cyan while s-curvature sources are shown in dark teal.}
\label{fig:c-s_curvature.png}
\end{figure}

Fig.~\ref{fig:c-s_curvature.png} shows the distribution of curved sources in $PD$ space. While most s-shaped sources are FRIIs, and most c-shaped sources are FRIs, this is not always the case. For example, any curvature in projection can be misidentified and some emission (such as those disturbed by a cluster) can exhibit characteristics of both c- and s-curvature. The fact that c-shaped sources appear at lower radio luminosities would be consistent with the idea that disturbances caused by motions with respect to the external medium are generally seen in sources with lower jet power.

\subsection{Precession indicators}
\label{sec:precession}
\begin{figure*}
\includegraphics[width=1\linewidth]{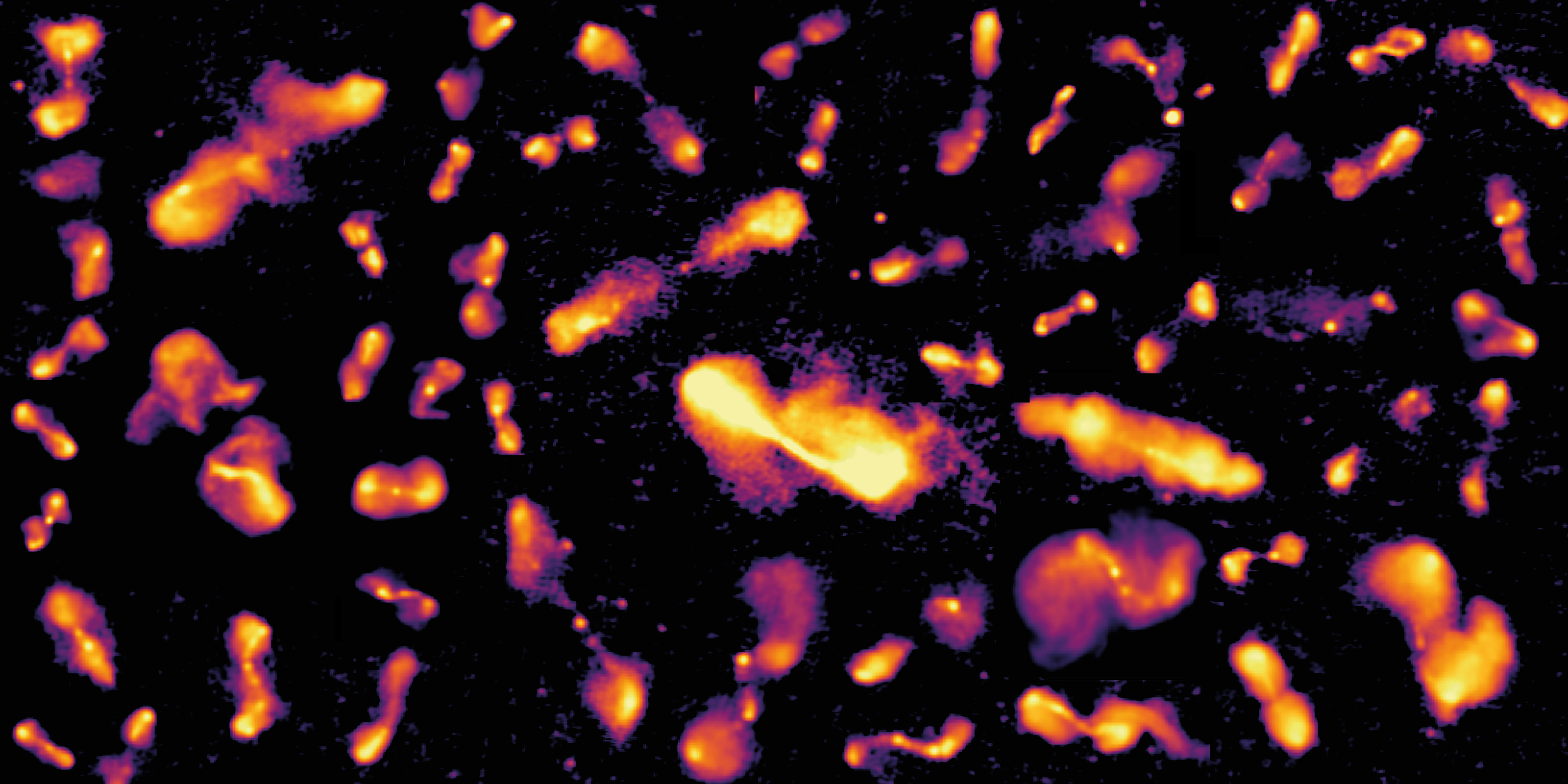}
\caption{Montage of representative sources tagged with three precession indicators. Details as in Fig.\ \ref{fig:hybrids.png}.}
\label{fig:precession.png}
\end{figure*}

\cite{krause18} identified four precession indicators: jet at edge of lobe (E), curvature in jet path (C), s-shaped symmetry (S) and multiple or complex hotspots (H). 

For this study, we do not define curvature (C) as a specific precession indicator since many curved sources are FRIs without any indication of precession, and precessing FRIIs can be captured just by the (S) indicator, which inherently requires the presence of curvature. As such, the three morphological indicators which are precession-specific are S-curvature, Misalignment and Multiple hotspots. These correspond to (S), (E) and (H) respectively in \cite{krause18}. 

We found that our classification system showed that 17\% of the full $2'$ flux limited catalogue showed evidence of a single precession indicator, 7\% showed any two and 5\% showed all three. Since these categories were mutually exclusive, around 29\% of sources in total show some indication of precession. This is likely to be an underrepresentation since it includes all FRIs, FRIIs, hybrids, relaxed doubles and restarts -- many of which are not expected to show signs of precession; or, perhaps more realistically, have not been examined to see if precession would produce different signatures in the population. One example of this is that multiple hotspots would not be a suitable precession indicator in FRI jets almost by definition, since FRI jets typically do not produce hotspots. This means that the current classification system systematically excludes FRIs from being defined as precessing, but it is certainly not the case that FRIs can never precess. 

In jet simulations, \cite{horton23} showed that some straight jets can produce multiple hotspots when no other indicators were present. This was particularly associated with young, rapidly-expanding lobes. Some of these sources were genuinely precessing and would later exhibit curvature and misalignment, while some were not precessing and showed different hotspot splitting mechanisms regardless. Given these simulation results, we wanted to look for any evidence of this in LoTSS DR2. The prevalence of precession indicators is summarised in Table \ref{tab:indicators}.

\begin{table}[]
    \centering
     \caption{Prevalence of precession indicators within the sample}
     \label{tab:indicators}
    \begin{tabular}{l | l | c | c }
         No. Indicators & Indicator & No. Sources & $z_{best}$ \\
         \hline 
         \multirow{3}{*}{One}
         & S-shaped symmetry (S) & 1082 & 864 \\
         & Misalignment (E) & 1287 & 1040 \\
         & Multiple Hotspots (M) & 2040 & 1613 \\
         & Any one & 2807 & 2220 \\
         \hline 
         \multirow{3}{*}{Two}
         & (S, E) & 842 & 681 \\
         & (M, S) & 532 & 434 \\
         & (E, M) & 692 & 561 \\
         & Any two & 1138 & 918 \\
         \hline
         \multirow{1}{*}{Three}
         & All & 464 & 379 \\
         \end{tabular}  
    \label{tab:precession}
    \end{table}

\begin{figure}
\includegraphics[width=1\linewidth]{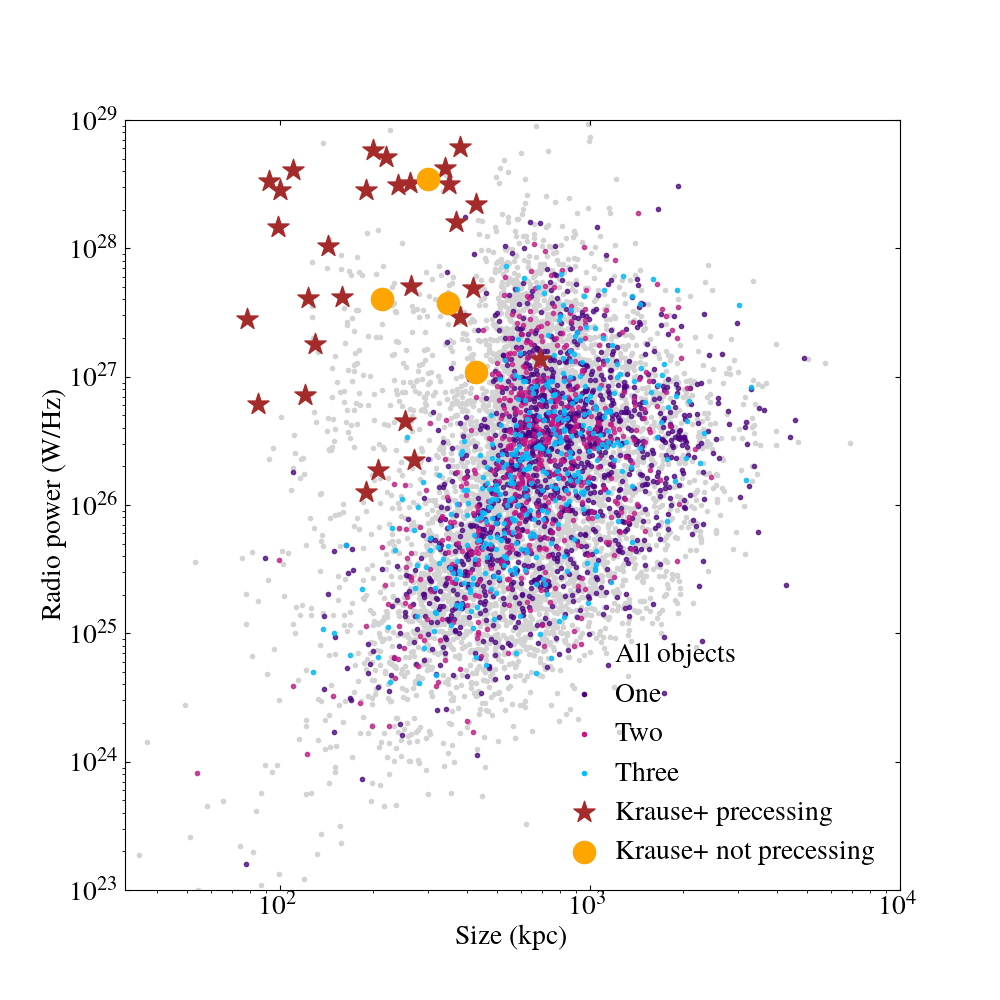}
\caption{$PD$ diagram showing the distribution of sources with any one (purple), two (magenta) and all three (pale blue) precession indicators. Overplotted are the positions of 3CRR sources examined by \cite{krause18}, where the red stars represent sources with potential precession indicators and the orange circles are those without.}
\label{fig:precession_dist.png}
\end{figure}

Fig.~\ref{fig:precession_dist.png} suggests that there are no size or luminosity constraints on the number of precession indicators, and if the prevalence of sources with all three (light blue) appears centralised, this is probably due to the smaller number of sources with all three. Indicator count is exclusive and any given source should only appear once. Relative to the work of \cite{krause18}, we find a smaller precession fraction overall but a much larger total number of potentially precessing sources. Our selection criteria mean that we sample a quite different population from \cite{krause18} with our sources being generally physically larger and lower in radio luminosity.

\subsection{Multiple hotspots in straight jets}

\cite{horton20b} showed that the prevalence of false negatives (precessing jets that show no features of precession) is much higher than that of false positives (precession indicators in jets that are not precessing). The most systematic occurrence of false positives was in the specific case of straight jets with multiple hotpots, as simulations have showed that these jets can frequently produce two concurrent hotspots early in the lifecycle of the jets \citep{horton23}. 

\begin{figure*}
\includegraphics[width=1\linewidth]{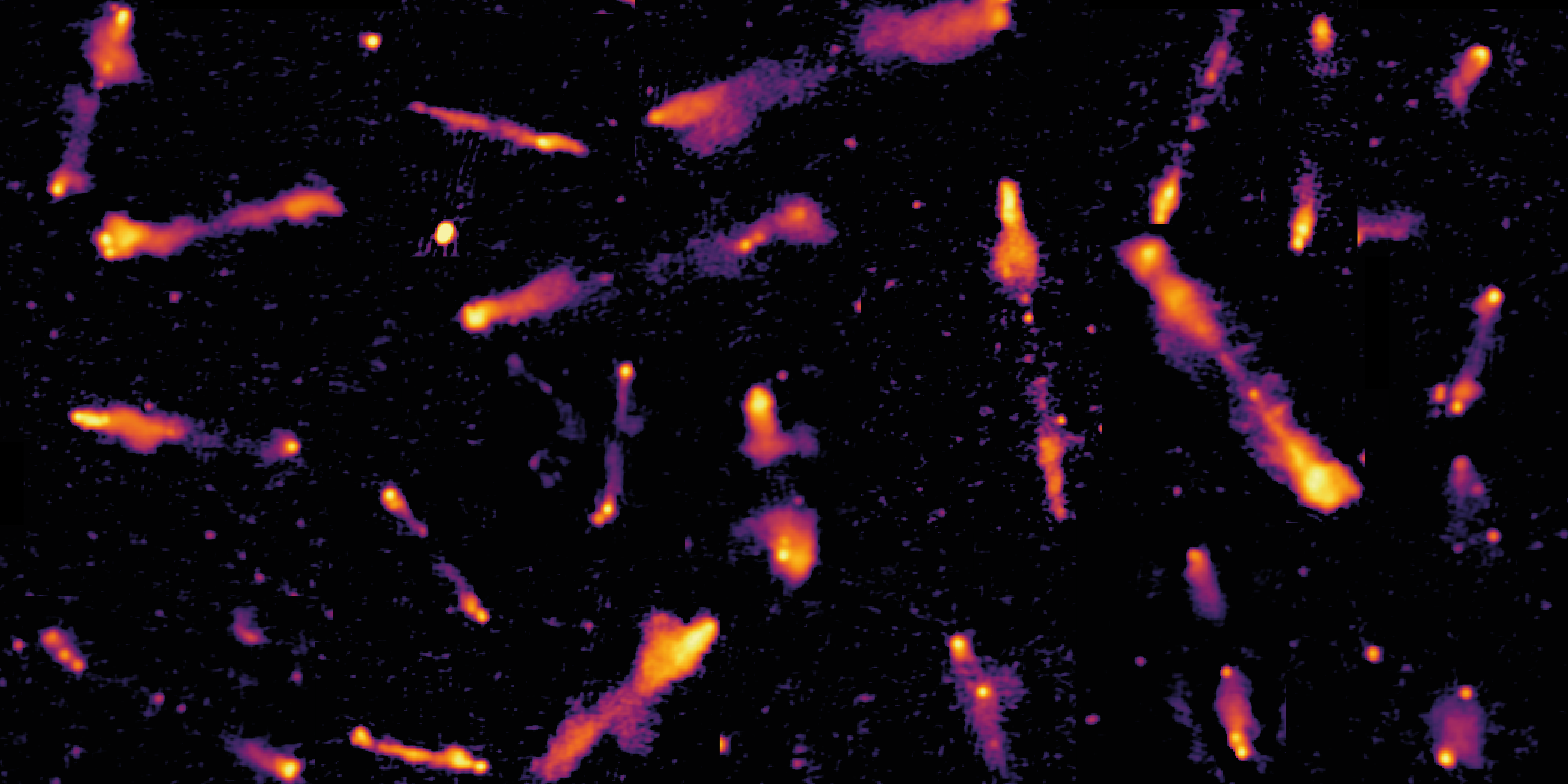}
\caption{Montage of representative sources tagged as having multiple or complex hotspots with apparently straight jets. Details as in Fig.\ \ref{fig:hybrids.png}.}
\label{fig:straight-multi.png}
\end{figure*}

Fig.~\ref{fig:straight-multi.png} shows a collection of objects which showed the possibility of having multiple hotspots during visual inspection. The classification system did nothing to define whether or not these were `true' hotspots in the sense of multiple terminal hotspots (see \cite{horton23}), jet knots, or shocked regions. And yet some objects, such as the second source in the top-left corner, do appear to distinctly show features that would be consistent with two terminal hotspots developing from a single jet. More work, and higher-resolution radio images, will be required to examine this population and more accurately classify the jet features to determine the prevalence of `true' multiple hotspots and their aetiology. 

\subsection{Precession indicators and mass}

\begin{figure}
\includegraphics[width=\linewidth]{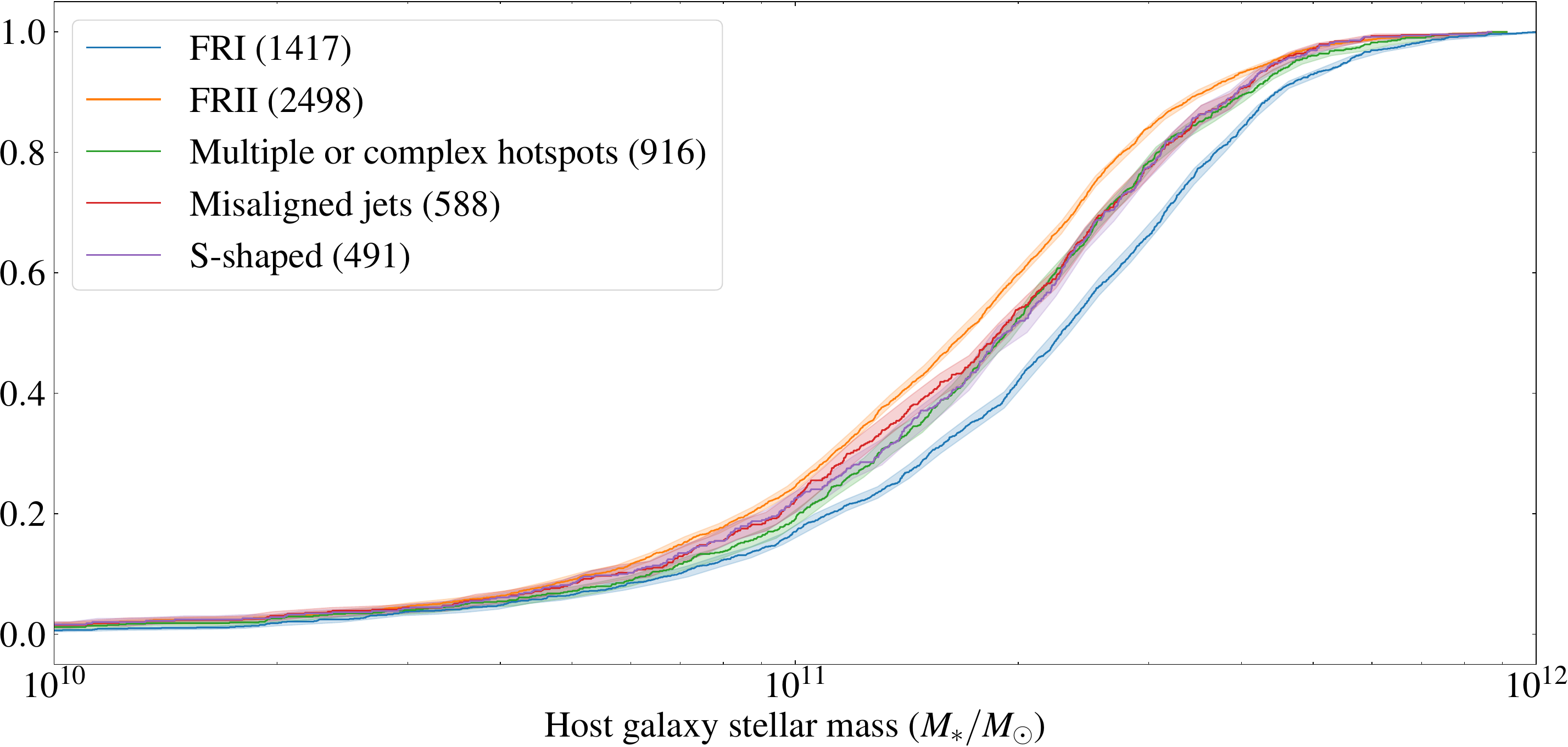}
\includegraphics[width=\linewidth]{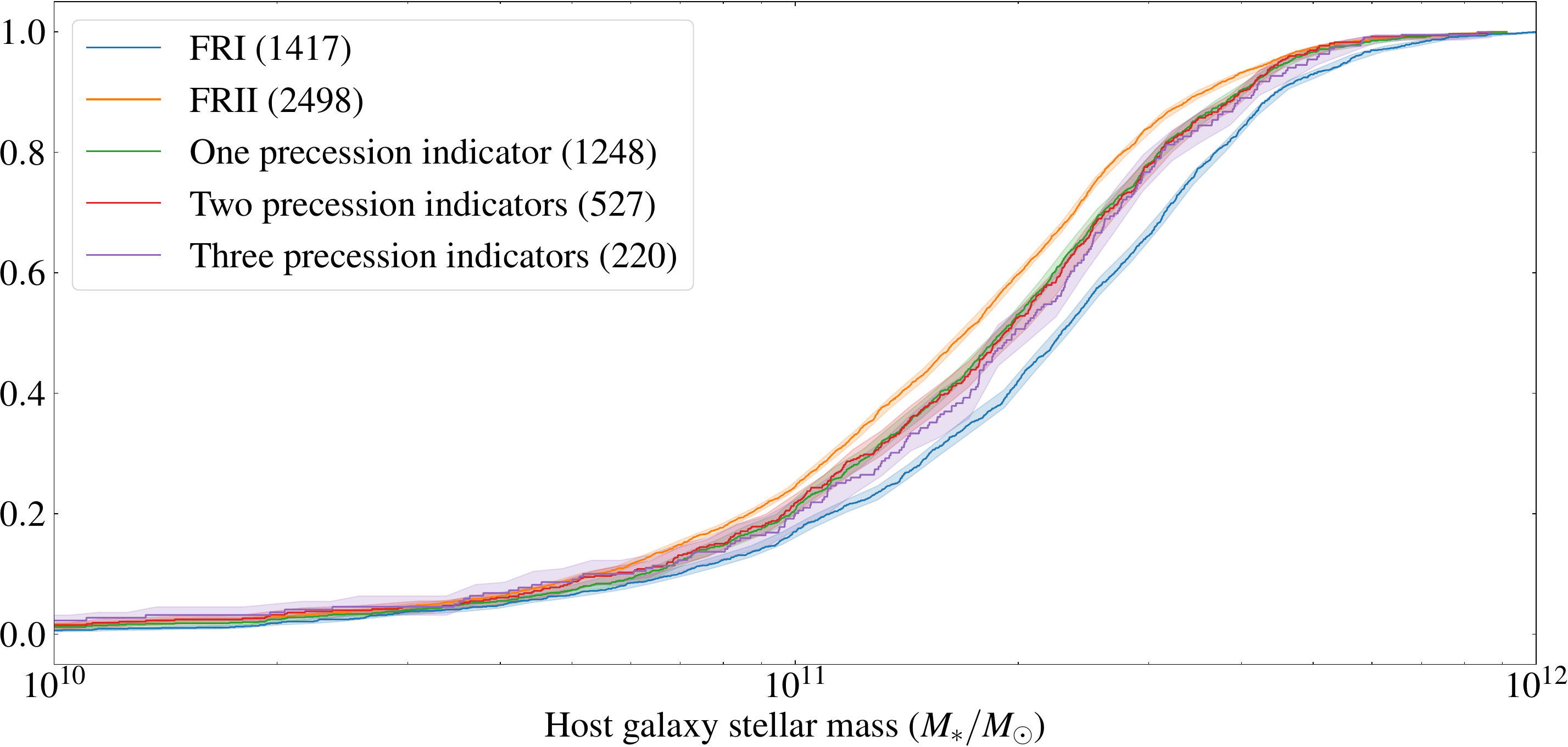}
\caption{Cumulative distribution of mass as a function of precession indicator. Top panel: individual indicators; bottom panel: combined indicators. Labels as in Fig.\ \ref{fig:massdist-mainclass}.}
\label{fig:massdist-precess}
\end{figure}

As discussed in Section \ref{sec:mass}, we have galaxy mass measurements for a large fraction of our sources, and so can relate the precession indicators we have measured to galaxy masses, with interesting results, as shown in Fig.\ \ref{fig:massdist-precess}. Objects with any of the precession indicators, or any combination of them, tend to have significantly more massive hosts than typical FRIIs, and this remains true even if we restrict our sample to only sources that are classified as FRIIs and also show these precession indicators. Given that our classifications are blind to the host galaxy mass, these results are a strong indication that they do pick out physically distinct populations. One possibility, of course, is that more massive galaxies have richer environments and so show more distortions of all types, but this suggestion fails to explain the relationship with multiple hotspots, which reflect fluid flow internal to the lobes. Alternatively, the relationship with mass may indicate that more massive galaxies are more likely to be experiencing the conditions required for detectable jet precession, such as more prevalent binary supermassive black holes.

\section{Discussion}
\subsection{Critique of methodology}
This work was reliant on subjective classification undertaken by visual inspection. The $PD$ diagrams and image collections shown in this paper suggest that many distinct populations can successfully be identified this way based on their morphological characteristics. However, it is very important to note the drawbacks and limitations of this approach. One critical issue is that many of the objects used in this classification were not well resolved, with only 15 beams across the source in many cases. For more distant objects, some elongated features and extended edge-brightened regions may be interpreted as being indicative of an underlying morphological feature, when perhaps those features might disappear at higher resolutions. Given the future availability of all-sky surveys at high resolution, we have chosen to keep broad selection criteria that could be available for further analysis rather than miss sources with interesting characteristics, but the tendency towards overestimation of the classifiability of an object must be seen as a key element in terms of reproducibility. 

It is also clear that a variety of other groupings and criteria could be used, both in terms of the initial classification and in creating subsets of distinct populations based on their collection of features. This approach was chosen as a direct consequence of the dual simulation and observations background of the authors; however, there is no guarantee that the same features would be chosen by any radio astronomer at any career stage. We encourage others to experiment with the catalogue and classification setup, which are released publicly with this paper; however, it is important to recognise that visual inspection of around 10,000 objects is a significant undertaking, and that presenting too many options to the classifier is likely to decrease the accuracy of the classification rather than increase it. 

Part of our initial classification setup predated discussions generated by \cite{rudnick21}, and so there is not necessarily deliberate and consistent overlap with other work on `tagging' radio galaxies. However, the morphological complexities shown in this work have allowed complex subclasses to emerge naturally even when the authors were not conscious of their existence. This suggests that language-based ordering of morphological features may indeed have a role to play in future classification studies; the work of \cite{bowles23} appears to support this through incorporating language-based algorithms, and it would be interesting to explore the value of that in future work. Additionally, work on cross-matching and optical IDs \citep{hardcastle23,mostert24,alegre24} shows increasing promise for improving component selection and galaxy classification.

While the role of machine learning is outside the scope of this current paper, we are aware that human-based classification would not have been possible for a much larger catalogue. Indeed, the need for certain categories only became clear partway through the classification, which led to difficulties with the `hybrid' class. Such problems would likely not exist within a pure ML context; and yet perhaps some of the relationships between sources would have been more difficult to detect in that framework.

\subsection{Prevalence of precession candidates}

We have shown that precession indicators are widespread in our sample, and while the fraction of potentially precessing objects that we report (29\% of all sources) is smaller than that discussed by \cite{krause18} (who reported that 73\% of their sources showed precession-related features), we have extended the detection of these indicators to larger physical sizes and lower luminosities (Fig.\ \ref{fig:precession_dist.png}). The presence of precession indicators appears to be related to the host galaxy mass.

It is important to note that, while the expectations from precession-driven hydrodynamics are well supported by simulations, few simulations are currently capable of differentiating between different sources of precession or accurately modelling the broad range of potential interactions in galactic centres that would influence jet-disk and black hole spin dynamics in the true sources we observe in this paper. Therefore, while it is possible that some -- even many -- of the 379 identifiable radio sources with three precession indicators could be candidate binary supermassive black holes, it is not possible to rule out any other causes of precession (or indeed other causes of systematic jet movement) without a far more robust analysis of the underlying populations. For example, complex dynamics of the host cluster environment \citep[e.g.][]{bourne21} could in principle generate some of the precession indicators that we see here. Moreover, while the simulations of \cite{horton20b} showed that false negatives are rare, it is very possible that hosts not identified in this study could be precessing. This is especially true of higher redshift, poorly resolved jets which are very difficult to interpret with current observations. Gentle or less extreme precession may be also be more difficult to detect. Overall, we could consider the fraction of sources with three clear precession indicators (5\%) to be a very conservative lower limit on the total fraction of sources exhibiting these indicators in the sky: its relation to the supermassive binary black hole population is less clear.

This fraction is significantly lower than the fraction of precession indicators reported for 3C objects by \cite{krause18}. Morphological classification is subjective by nature and some discrepancy between the results can be expected: \cite{krause18} describe their results as giving an upper limit on the precession fraction. But there are also possible physical reasons for the difference. There are many possible reasons for this discrepancy: (1) 3C objects could be unusual for their given redshift, and so by selection are more luminous and exhibit more extreme characteristics; (2) that the Krause sample is dominated by FRII-type sources, where we also see high-precession indicator sources; (3) Perhaps the most important factor is that 3C objects are very well-studied and are associated with many high resolution observations at a range of frequencies and as such it is easier to identify features compared to lower resolution survey data; (4) it could also be that the subjective nature of what counts as curvature or edgeness, or indeed peculiarity, makes it difficult to systematically quantify what is or is not a precession indicator. As we have shown previously \citep{horton20b}, apparently straight jets with no precession indicators can be produced even from a precessing system, particularly early on in a source's lifetime.

\subsection{Implications for variability in jet power} 

The morphology of LoTSS extended radio sources traces the history of the jet input over the radio source lifetimes, i.e. timescales of  $\sim$ $10^7 -10^9$ y. Studying the source structures therefore provides information on the variability and life cycles of the nuclear AGN hosts. Many sources associated with our `relaxed' class of objects exhibit large physical sizes and relatively low luminosity (Fig. \ref{fig:scatter_fris_friis.png} and Fig.\ \ref{fig:scatter_relaxed.png}). These are the expected characteristics of remnant radio galaxies, which will fade rapidly once the jet is switched off \citep{slee01,hardcastle18,mahatma18,jurlin21,mostert23}. Their location at lower luminosities than, but similar physical sizes to, the FRII population suggests that they may have evolved from an initial active FRII phase, consistent with their mostly well-bounded morphology (Fig.\ \ref{fig:relaxed.png}) and a mass distribution consistent with that of the FRII population (Fig.\ \ref{fig:massdist-mainclass}).  The fraction of relaxed sources is small ($\sim 5$\% of the total), implying that radio lobes fade relatively quickly after the jet switches off.

Objects classed as `restarted', interpreted as extreme cases of jet variability, also lie in a distinct region of the $PD$ diagram, with similar power to but larger physical size than the bulk of the FRIIs. Physically these sources are expected to be the result of an interruption in the supply of energy to the jet. The jet must then have restarted with roughly comparable power after a relatively short interruption (if the jet is much less powerful, we will not observe a restarted inner lobe, while if it is much more powerful, the emission from the new lobe will dominate the old one; if the interruption is shorter than the light travel time to the end of the lobe, there will be little observable effect, while if it is too long the outer lobes will fade below the limit of detectability). The somewhat larger physical size of the detected restarts is then expected physically, as these can only be detected in relatively old sources, but it is also possible that our classification is biased against small remnants where we might not have the resolution to detect a new pair of inner lobes. Although the sample size of restarted objects is small ($\sim 2$\% of the total), the mass distribution suggests that they may be associated with slightly more massive galaxies. Confirmation of such a trend and studies of larger samples of restarted sources at higher resolutions and over larger redshift ranges  could potentially yield important information about merger rate and galaxy evolution. 

\subsection{Scope for future work}
Many of these classifications require follow-up observations at higher resolution, and would benefit from an exploration of widefield dynamics, spectral properties, galaxy mass and potential environmental interactions. 

The coming era of large astronomical surveys and increased sensitivity will allow for numerous avenues to be explored. This specific population is likely to be studied in much more detail due to the provisions of WEAVE-LOFAR \citep{smith16}, LoTSS DR3, LOFAR 2.0 and any relevant deep field studies that overlap with any of these sources. Additionally, similar studies are likely to be available for the Southern sky and equatorial regions through ASKAP EMU, MeerKAT and, eventually, the SKA. LOFAR data in particular will benefit from the routine exploitation of the LOFAR long baselines in the coming years \citep{morabito22} which would allow us to greatly improve our selection of precession indicators. 

It is worth remembering that the known angular size redshift relation shows that few $z>2$ radio sources have angular sizes larger than $15''$, so the scope of this morphological study has inevitably been limited in redshift by the $6''$ resolution of LoTSS: indeed, only 40 such high-$z$ sources exist in our sample with only three showing one or more precession indicators. Future LOFAR large-sky surveys using the international baselines will reach sub-arcsecond resolutions will open up the exciting prospect of carrying out similar robust statistical analyses of how the morphological properties of radio sources and its relation with their optical hosts vary as a function of redshift. Such surveys will use the unique diagnostics of radio morphologies to study the role of radio AGN/SMBHs on the formation and evolution of galaxies and as probes of the high-redshift medium.

One particularly exciting possibility is the eventual combination of the catalogue of precession-indicated sources with future gravitational and continuous wave studies in the supermassive regime \citep[e.g.,][]{mingarelli25,steinle24,buonanno14}. Understanding the influence of potential precession sources on the range of mass effects observed in this population is beyond the scope of this paper, but our catalogue provides a starting point for understanding the parameter space between precession mechanisms, precession indicators, hypothetical supermassive binary separation, accretion disk dynamics and bulk environmental factors.

\subsection{Summary of key results}
This paper explores a large population of hand-selected radio galaxies. Our main findings and results for future exploration are summarised below:

\begin{itemize}
    \item A population of galaxies with multiple precession indicators which might be the most likely candidate for host supermassive binary black hole systems;
    \item Some evidence that host galaxy mass appears to scale slightly with the number of precession indicators;
    \item A distinct population of galaxies exhibiting features of both FRI and FRII galaxies (with or without precession characteristics), where size, luminosity and galaxy mass are intermediate to FRI and FRII populations;
    \item A potentially new population of point-like sources which show highly unusual characteristics, but which may or may not be explained by deconvolution effects;
    \item Confirmation of many similar surveys regarding prevalence of x-shaped and z-shaped radio galaxies and peculiar morphology in general.
\end{itemize}

\subsection{Final conclusions}
Simulations have long identified certain morphological characteristics as related to jet precession. Through a process of visual inspection and manual tagging of large and bright radio sources, we have found that complex morphologies are common in this population. By assigning non-exclusive labels to sources we can see that distinct population trends emerge for some features. For example, jets that could not easily be described as ``FRI'' or ``FRII'' based solely on morphology tended to occupy $PD$ diagrammatic space between more identifiable FRIs and FRIIs. ``Relaxed doubles,'' which are assumed to come from host AGN that have been turned ``off'', tended to have larger size and lower luminosity than active galaxies, which would be consistent with adiabatic expansion of old lobe material. 

We have seen a subpopulation of jets with different indicators of precession, and these show particular characteristics in terms of size, mass and luminosity alongside distinctive morphological features. By inspecting large numbers of sources from the LOFAR survey we have greatly increased the number and diversity of candidate precessing jet sources.

The primary drawbacks of this study include the subjective nature of the morphological classification; difficulties interpreting images due to poor resolution (common), interference from imaging artefacts (rare), and difficulty clearly identifying features (common in some populations); the relatively small field of view surrounding many sources which meant some crucial elements (e.g., nearby clusters) were often missed; lack of appropriate knowledge in classifying certain sources (such as for the population of bright, small objects discussed in section~\ref{subsec:pointsources}.

Overall, the catalogue as a whole provides a rich dataset for exploring complex morphologies and their relationship to different stages of galactic evolution, but leaves open many questions regarding the morphology and evolution of large source samples.

\begin{acknowledgements}
The authors would like to thank the anonymous reviewer for significantly improving this paper, and Arpita Misra for helpful comments on an earlier draft.  

MAH acknowledges the support of the UK STFC under a grant to the University of Cambridge [ST/Y000447/1], and prior support under STFC grants [ST/R504786/1, ST/R000905/1, ST/X002543/1]. MJH thanks the UK STFC for support [ST/V000624/1,  ST/Y001249/1].

LOFAR is the Low Frequency Array, designed and constructed by ASTRON. It has observing, data processing, and data storage facilities in several countries, which are owned by various parties (each with their own funding sources), and which are collectively operated by the ILT foundation under a joint scientific policy. The ILT resources have benefited from the following recent major funding sources: CNRS-INSU, Observatoire de Paris and Université d'Orléans, France; BMBF, MIWF-NRW, MPG, Germany; Science Foundation Ireland (SFI), Department of Business, Enterprise and Innovation (DBEI), Ireland; NWO, The Netherlands; The Science and Technology Facilities Council, UK; Ministry of Science and Higher Education, Poland; The Istituto Nazionale di Astrofisica (INAF), Italy.

This research made use of the Dutch national e-infrastructure with support of the SURF Cooperative (e-infra 180169) and the LOFAR e-infra group. The Jülich LOFAR Long Term Archive and the German LOFAR network are both coordinated and operated by the Jülich Supercomputing Centre (JSC), and computing resources on the supercomputer JUWELS at JSC were provided by the Gauss Centre for Supercomputing e.V. (grant CHTB00) through the John von Neumann Institute for Computing (NIC).

This research made use of the University of Hertfordshire
high-performance computing facility and the LOFAR-UK computing
facility located at the University of Hertfordshire (\url{https://uhhpc.herts.ac.uk}) and supported by
STFC [ST/P000096/1], and of the Italian LOFAR IT computing
infrastructure supported and operated by INAF, and by the Physics
Department of Turin University (under an agreement with Consorzio
Interuniversitario per la Fisica Spaziale) at the C3S Supercomputing
Centre, Italy.

This research made use of {\sc Astropy}, a
community-developed core Python package for astronomy
\citep{AstropyCollaboration13} hosted at
\url{http://www.astropy.org/}, of {\sc Matplotlib} \citep{Hunter07},
of {\sc APLpy}, an open-source astronomical plotting package for
Python hosted at \url{http://aplpy.github.com/}, and of {\sc topcat}
and {\sc stilts} \citep{Taylor05}.
\end{acknowledgements}
  
\bibliographystyle{aa}
\renewcommand{\refname}{REFERENCES}
\bibliography{horton}

\begin{thebibliography}{89}
\expandafter\ifx\csname natexlab\endcsname\relax\def\natexlab#1{#1}\fi

\bibitem[{{Agazie} {et~al.}(2023{\natexlab{a}}){Agazie}, {Anumarlapudi},
  {Archibald}, {Arzoumanian}, {Baker}, {B{\'e}csy}, {Blecha}, {Brazier},
  {Brook}, {Burke-Spolaor}, {Burnette}, {Case}, {Charisi}, {Chatterjee},
  {Chatziioannou}, {Cheeseboro}, {Chen}, {Cohen}, {Cordes}, {Cornish},
  {Crawford}, {Cromartie}, {Crowter}, {Cutler}, {Decesar}, {Degan}, {Demorest},
  {Deng}, {Dolch}, {Drachler}, {Ellis}, {Ferrara}, {Fiore}, {Fonseca},
  {Freedman}, {Garver-Daniels}, {Gentile}, {Gersbach}, {Glaser}, {Good},
  {G{\"u}ltekin}, {Hazboun}, {Hourihane}, {Islo}, {Jennings}, {Johnson},
  {Jones}, {Kaiser}, {Kaplan}, {Kelley}, {Kerr}, {Key}, {Klein}, {Laal}, {Lam},
  {Lamb}, {Lazio}, {Lewandowska}, {Littenberg}, {Liu}, {Lommen}, {Lorimer},
  {Luo}, {Lynch}, {Ma}, {Madison}, {Mattson}, {McEwen}, {McKee}, {McLaughlin},
  {McMann}, {Meyers}, {Meyers}, {Mingarelli}, {Mitridate}, {Natarajan}, {Ng},
  {Nice}, {Ocker}, {Olum}, {Pennucci}, {Perera}, {Petrov}, {Pol}, {Radovan},
  {Ransom}, {Ray}, {Romano}, {Sardesai}, {Schmiedekamp}, {Schmiedekamp},
  {Schmitz}, {Schult}, {Shapiro-Albert}, {Siemens}, {Simon}, {Siwek}, {Stairs},
  {Stinebring}, {Stovall}, {Sun}, {Susobhanan}, {Swiggum}, {Taylor}, {Taylor},
  {Turner}, {Unal}, {Vallisneri}, {van Haasteren}, {Vigeland}, {Wahl}, {Wang},
  {Witt}, {Young}, \& {Nanograv Collaboration}}]{agazie23}
{Agazie}, G., {Anumarlapudi}, A., {Archibald}, A.~M., {et~al.}
  2023{\natexlab{a}}, \apjl, 951, L8

\bibitem[{{Agazie} {et~al.}(2023{\natexlab{b}}){Agazie}, {Anumarlapudi},
  {Archibald}, {Arzoumanian}, {Baker}, {B{\'e}csy}, {Blecha}, {Brazier},
  {Brook}, {Burke-Spolaor}, {Case}, {Casey-Clyde}, {Charisi}, {Chatterjee},
  {Cohen}, {Cordes}, {Cornish}, {Crawford}, {Cromartie}, {Crowter}, {Decesar},
  {Demorest}, {Digman}, {Dolch}, {Drachler}, {Ferrara}, {Fiore}, {Fonseca},
  {Freedman}, {Garver-Daniels}, {Gentile}, {Glaser}, {Good}, {G{\"u}ltekin},
  {Hazboun}, {Hourihane}, {Jennings}, {Johnson}, {Jones}, {Kaiser}, {Kaplan},
  {Kelley}, {Kerr}, {Key}, {Laal}, {Lam}, {Lamb}, {Lazio}, {Lewandowska},
  {Liu}, {Lorimer}, {Luo}, {Lynch}, {Ma}, {Madison}, {McEwen}, {McKee},
  {McLaughlin}, {McMann}, {Meyers}, {Meyers}, {Mingarelli}, {Mitridate}, {Ng},
  {Nice}, {Ocker}, {Olum}, {Pennucci}, {Perera}, {Petrov}, {Pol}, {Radovan},
  {Ransom}, {Ray}, {Romano}, {Sardesai}, {Schmiedekamp}, {Schmiedekamp},
  {Schmitz}, {Shapiro-Albert}, {Siemens}, {Simon}, {Siwek}, {Stairs},
  {Stinebring}, {Stovall}, {Susobhanan}, {Swiggum}, {Taylor}, {Taylor},
  {Turner}, {Unal}, {Vallisneri}, {van Haasteren}, {Vigeland}, {Wahl}, {Witt},
  {Young}, \& {Nanograv Collaboration}}]{agazie23b}
{Agazie}, G., {Anumarlapudi}, A., {Archibald}, A.~M., {et~al.}
  2023{\natexlab{b}}, \apjl, 951, L50

\bibitem[{{Alegre} {et~al.}(2024){Alegre}, {Best}, {Sabater}, {R{\"o}ttgering},
  {Hardcastle}, \& {Williams}}]{alegre24}
{Alegre}, L., {Best}, P., {Sabater}, J., {et~al.} 2024, \mnras, 532, 3322

\bibitem[{{Arzoumanian} {et~al.}(2020){Arzoumanian}, {Baker}, {Blumer},
  {B{\'e}csy}, {Brazier}, {Brook}, {Burke-Spolaor}, {Chatterjee}, {Chen},
  {Cordes}, {Cornish}, {Crawford}, {Cromartie}, {Decesar}, {Demorest}, {Dolch},
  {Ellis}, {Ferrara}, {Fiore}, {Fonseca}, {Garver-Daniels}, {Gentile}, {Good},
  {Hazboun}, {Holgado}, {Islo}, {Jennings}, {Jones}, {Kaiser}, {Kaplan},
  {Kelley}, {Key}, {Laal}, {Lam}, {Lazio}, {Lorimer}, {Luo}, {Lynch},
  {Madison}, {McLaughlin}, {Mingarelli}, {Ng}, {Nice}, {Pennucci}, {Pol},
  {Ransom}, {Ray}, {Shapiro-Albert}, {Siemens}, {Simon}, {Spiewak}, {Stairs},
  {Stinebring}, {Stovall}, {Sun}, {Swiggum}, {Taylor}, {Turner}, {Vallisneri},
  {Vigeland}, {Witt}, \& {Nanograv Collaboration}}]{arzoumanian20}
{Arzoumanian}, Z., {Baker}, P.~T., {Blumer}, H., {et~al.} 2020, \apjl, 905, L34

\bibitem[{{Astropy Collaboration} {et~al.}(2013){Astropy Collaboration},
  {Robitaille}, {Tollerud}, {Greenfield}, {Droettboom}, {Bray}, {Aldcroft},
  {Davis}, {Ginsburg}, {Price-Whelan}, {Kerzendorf}, {Conley}, {Crighton},
  {Barbary}, {Muna}, {Ferguson}, {Grollier}, {Parikh}, {Nair}, {Unther},
  {Deil}, {Woillez}, {Conseil}, {Kramer}, {Turner}, {Singer}, {Fox}, {Weaver},
  {Zabalza}, {Edwards}, {Azalee Bostroem}, {Burke}, {Casey}, {Crawford},
  {Dencheva}, {Ely}, {Jenness}, {Labrie}, {Lim}, {Pierfederici}, {Pontzen},
  {Ptak}, {Refsdal}, {Servillat}, \& {Streicher}}]{AstropyCollaboration13}
{Astropy Collaboration}, {Robitaille}, T.~P., {Tollerud}, E.~J., {et~al.} 2013,
  \aap, 558, A33

\bibitem[{{Babak} {et~al.}(2016){Babak}, {Petiteau}, {Sesana}, {Brem},
  {Rosado}, {Taylor}, {Lassus}, {Hessels}, {Bassa}, {Burgay}, {Caballero},
  {Champion}, {Cognard}, {Desvignes}, {Gair}, {Guillemot}, {Janssen},
  {Karuppusamy}, {Kramer}, {Lazarus}, {Lee}, {Lentati}, {Liu}, {Mingarelli},
  {Os{\l}owski}, {Perrodin}, {Possenti}, {Purver}, {Sanidas}, {Smits},
  {Stappers}, {Theureau}, {Tiburzi}, {van Haasteren}, {Vecchio}, \&
  {Verbiest}}]{babak16}
{Babak}, S., {Petiteau}, A., {Sesana}, A., {et~al.} 2016, \mnras, 455, 1665

\bibitem[{{Babak} \& {Sesana}(2012)}]{babak12}
{Babak}, S. \& {Sesana}, A. 2012, \prd, 85, 044034

\bibitem[{{Baldi} {et~al.}(2016){Baldi}, {Capetti}, \& {Giovannini}}]{baldi16}
{Baldi}, R.~D., {Capetti}, A., \& {Giovannini}, G. 2016, Astronomische
  Nachrichten, 337, 114

\bibitem[{{Barthel} {et~al.}(1988){Barthel}, {Miley}, {Schilizzi}, \&
  {Lonsdale}}]{barthel88}
{Barthel}, P.~D., {Miley}, G.~K., {Schilizzi}, R.~T., \& {Lonsdale}, C.~J.
  1988, \aaps, 73, 515

\bibitem[{{Becker} {et~al.}(1995){Becker}, {White}, \& {Helfand}}]{becker95}
{Becker}, R.~H., {White}, R.~L., \& {Helfand}, D.~J. 1995, \apj, 450, 559

\bibitem[{{Blanton} {et~al.}(2017){Blanton}, {Bershady}, {Abolfathi},
  {Albareti}, {Allende Prieto}, {Almeida}, {Alonso-Garc{\'\i}a}, {Anders},
  {Anderson}, {Andrews}, {Aquino-Ort{\'\i}z}, {Arag{\'o}n-Salamanca},
  {Argudo-Fern{\'a}ndez}, {Armengaud}, {Aubourg}, {Avila-Reese}, {Badenes},
  {Bailey}, {Barger}, {Barrera-Ballesteros}, {Bartosz}, {Bates}, {Baumgarten},
  {Bautista}, {Beaton}, {Beers}, {Belfiore}, {Bender}, {Berlind}, {Bernardi},
  {Beutler}, {Bird}, {Bizyaev}, {Blanc}, {Blomqvist}, {Bolton}, {Boquien},
  {Borissova}, {van den Bosch}, {Bovy}, {Brandt}, {Brinkmann}, {Brownstein},
  {Bundy}, {Burgasser}, {Burtin}, {Busca}, {Cappellari}, {Delgado Carigi},
  {Carlberg}, {Carnero Rosell}, {Carrera}, {Chanover}, {Cherinka}, {Cheung},
  {G{\'o}mez Maqueo Chew}, {Chiappini}, {Choi}, {Chojnowski}, {Chuang},
  {Chung}, {Cirolini}, {Clerc}, {Cohen}, {Comparat}, {da Costa}, {Cousinou},
  {Covey}, {Crane}, {Croft}, {Cruz-Gonzalez}, {Garrido Cuadra}, {Cunha},
  {Damke}, {Darling}, {Davies}, {Dawson}, {de la Macorra}, {Dell'Agli}, {De
  Lee}, {Delubac}, {Di Mille}, {Diamond-Stanic}, {Cano-D{\'\i}az}, {Donor},
  {Downes}, {Drory}, {du Mas des Bourboux}, {Duckworth}, {Dwelly}, {Dyer},
  {Ebelke}, {Eigenbrot}, {Eisenstein}, {Emsellem}, {Eracleous}, {Escoffier},
  {Evans}, {Fan}, {Fern{\'a}ndez-Alvar}, {Fernandez-Trincado}, {Feuillet},
  {Finoguenov}, {Fleming}, {Font-Ribera}, {Fredrickson}, {Freischlad},
  {Frinchaboy}, {Fuentes}, {Galbany}, {Garcia-Dias},
  {Garc{\'\i}a-Hern{\'a}ndez}, {Gaulme}, {Geisler}, {Gelfand},
  {Gil-Mar{\'\i}n}, {Gillespie}, {Goddard}, {Gonzalez-Perez}, {Grabowski},
  {Green}, {Grier}, {Gunn}, {Guo}, {Guy}, {Hagen}, {Hahn}, {Hall}, {Harding},
  {Hasselquist}, {Hawley}, {Hearty}, {Gonzalez Hern{\'a}ndez}, {Ho}, {Hogg},
  {Holley-Bockelmann}, {Holtzman}, {Holzer}, {Huehnerhoff}, {Hutchinson},
  {Hwang}, {Ibarra-Medel}, {da Silva Ilha}, {Ivans}, {Ivory}, {Jackson},
  {Jensen}, {Johnson}, {Jones}, {J{\"o}nsson}, {Jullo}, {Kamble}, {Kinemuchi},
  {Kirkby}, {Kitaura}, {Klaene}, {Knapp}, {Kneib}, {Kollmeier}, {Lacerna},
  {Lane}, {Lang}, {Law}, {Lazarz}, {Lee}, {Le Goff}, {Liang}, {Li}, {Li},
  {Lian}, {Lima}, {Lin}, {Lin}, {Bertran de Lis}, {Liu}, {de Icaza Lizaola},
  {Long}, {Lucatello}, {Lundgren}, {MacDonald}, {Deconto Machado}, {MacLeod},
  {Mahadevan}, {Geimba Maia}, {Maiolino}, {Majewski}, {Malanushenko},
  {Malanushenko}, {Manchado}, {Mao}, {Maraston}, {Marques-Chaves}, {Masseron},
  {Masters}, {McBride}, {McDermid}, {McGrath}, {McGreer}, {Medina Pe{\~n}a},
  {Melendez}, {Merloni}, {Merrifield}, {Meszaros}, {Meza}, {Minchev},
  {Minniti}, {Miyaji}, {More}, {Mulchaey}, {M{\"u}ller-S{\'a}nchez}, {Muna},
  {Munoz}, {Myers}, {Nair}, {Nandra}, {Correa do Nascimento}, {Negrete},
  {Ness}, {Newman}, {Nichol}, {Nidever}, {Nitschelm}, {Ntelis}, {O'Connell},
  {Oelkers}, {Oravetz}, {Oravetz}, {Pace}, {Padilla}, {Palanque-Delabrouille},
  {Alonso Palicio}, {Pan}, {Parejko}, {Parikh}, {P{\^a}ris}, {Park}, {Patten},
  {Peirani}, {Pellejero-Ibanez}, {Penny}, {Percival}, {Perez-Fournon},
  {Petitjean}, {Pieri}, {Pinsonneault}, {Pisani}, {Poleski}, {Prada},
  {Prakash}, {Queiroz}, {Raddick}, {Raichoor}, {Barboza Rembold}, {Richstein},
  {Riffel}, {Riffel}, {Rix}, {Robin}, {Rockosi}, {Rodr{\'\i}guez-Torres},
  {Roman-Lopes}, {Rom{\'a}n-Z{\'u}{\~n}iga}, {Rosado}, {Ross}, {Rossi}, {Ruan},
  {Ruggeri}, {Rykoff}, {Salazar-Albornoz}, {Salvato}, {S{\'a}nchez}, {Aguado},
  {S{\'a}nchez-Gallego}, {Santana}, {Santiago}, {Sayres}, {Schiavon}, {da Silva
  Schimoia}, {Schlafly}, {Schlegel}, {Schneider}, {Schultheis}, {Schuster},
  {Schwope}, {Seo}, {Shao}, {Shen}, {Shetrone}, {Shull}, {Simon}, {Skinner},
  {Skrutskie}, {Slosar}, {Smith}, {Sobeck}, {Sobreira}, {Somers}, {Souto},
  {Stark}, {Stassun}, {Stauffer}, {Steinmetz}, {Storchi-Bergmann},
  {Streblyanska}, {Stringfellow}, {Su{\'a}rez}, {Sun}, {Suzuki}, {Szigeti},
  {Taghizadeh-Popp}, {Tang}, {Tao}, {Tayar}, {Tembe}, {Teske}, {Thakar},
  {Thomas}, {Thompson}, {Tinker}, {Tissera}, {Tojeiro}, {Hernandez Toledo}, {de
  la Torre}, {Tremonti}, {Troup}, {Valenzuela}, {Martinez Valpuesta},
  {Vargas-Gonz{\'a}lez}, {Vargas-Maga{\~n}a}, {Vazquez}, {Villanova}, {Vivek},
  {Vogt}, {Wake}, {Walterbos}, {Wang}, {Weaver}, {Weijmans}, {Weinberg},
  {Westfall}, {Whelan}, {Wild}, {Wilson}, {Wood-Vasey}, {Wylezalek}, {Xiao},
  {Yan}, {Yang}, {Ybarra}, {Y{\`e}che}, {Zakamska}, {Zamora}, {Zarrouk},
  {Zasowski}, {Zhang}, {Zhao}, {Zheng}, {Zheng}, {Zhou}, {Zhou}, {Zhu},
  {Zoccali}, \& {Zou}}]{blanton17}
{Blanton}, M.~R., {Bershady}, M.~A., {Abolfathi}, B., {et~al.} 2017, \aj, 154,
  28

\bibitem[{{Bourne} {et~al.}(2024){Bourne}, {Fiacconi}, {Sijacki}, {Piotrowska},
  \& {Koudmani}}]{bourne24}
{Bourne}, M.~A., {Fiacconi}, D., {Sijacki}, D., {Piotrowska}, J.~M., \&
  {Koudmani}, S. 2024, \mnras, 534, 3448

\bibitem[{{Bourne} \& {Sijacki}(2021)}]{bourne21}
{Bourne}, M.~A. \& {Sijacki}, D. 2021, \mnras, 506, 488

\bibitem[{{Bowles} {et~al.}(2023){Bowles}, {Tang}, {Vardoulaki}, {Alexander},
  {Luo}, {Rudnick}, {Walmsley}, {Porter}, {Scaife}, {Slijepcevic}, {Adams},
  {Drabent}, {Dugdale}, {G{\"u}rkan}, {Hopkins}, {Jimenez-Andrade}, {Leahy},
  {Norris}, {Rahman}, {Ouyang}, {Segal}, {Shabala}, \& {Wong}}]{bowles23}
{Bowles}, M., {Tang}, H., {Vardoulaki}, E., {et~al.} 2023, \mnras, 522, 2584

\bibitem[{{Bridle} \& {Perley}(1984)}]{Bridle+Perley84}
{Bridle}, A.~H. \& {Perley}, R.~A. 1984, \araa, 22, 319

\bibitem[{{Brienza} {et~al.}(2017){Brienza}, {Godfrey}, {Morganti}, {Prandoni},
  {Harwood}, {Mahony}, {Hardcastle}, {Murgia}, {R{\"o}ttgering}, {Shimwell}, \&
  {Shulevski}}]{Brienza+17}
{Brienza}, M., {Godfrey}, L., {Morganti}, R., {et~al.} 2017, \aap, 606, A98

\bibitem[{{Buonanno} \& {Sathyaprakash}(2014)}]{buonanno14}
{Buonanno}, A. \& {Sathyaprakash}, B.~S. 2014, arXiv e-prints, arXiv:1410.7832

\bibitem[{{Burke-Spolaor}(2011)}]{burkespolaor11}
{Burke-Spolaor}, S. 2011, \mnras, 410, 2113

\bibitem[{{Caproni} {et~al.}(2007){Caproni}, {Abraham}, {Livio}, \& {Mosquera
  Cuesta}}]{caproni07}
{Caproni}, A., {Abraham}, Z., {Livio}, M., \& {Mosquera Cuesta}, H.~J. 2007,
  \mnras, 379, 135

\bibitem[{{Cheung}(2007)}]{cheung07}
{Cheung}, C.~C. 2007, \aj, 133, 2097

\bibitem[{{Condon} {et~al.}(1998){Condon}, {Cotton}, {Greisen}, {Yin},
  {Perley}, {Taylor}, \& {Broderick}}]{condon98}
{Condon}, J.~J., {Cotton}, W.~D., {Greisen}, E.~W., {et~al.} 1998, \aj, 115,
  1693

\bibitem[{{Cotton} {et~al.}(2020){Cotton}, {Thorat}, {Condon}, {Frank},
  {J{\'o}zsa}, {White}, {Deane}, {Oozeer}, {Atemkeng}, {Bester}, {Fanaroff},
  {Kupa}, {Smirnov}, {Mauch}, {Krishnan}, \& {Camilo}}]{cotton20}
{Cotton}, W.~D., {Thorat}, K., {Condon}, J.~J., {et~al.} 2020, \mnras
  [\eprint[arXiv]{2005.02723}]

\bibitem[{{Davis} \& {Tchekhovskoy}(2020)}]{davis20}
{Davis}, S.~W. \& {Tchekhovskoy}, A. 2020, \araa, 58, 407

\bibitem[{{de Gasperin} {et~al.}(2015){de Gasperin}, {Ogrean}, {van Weeren},
  {Dawson}, {Br{\"u}ggen}, {Bonafede}, \& {Simionescu}}]{degasperin15}
{de Gasperin}, F., {Ogrean}, G.~A., {van Weeren}, R.~J., {et~al.} 2015, \mnras,
  448, 2197

\bibitem[{{Dennett-Thorpe} {et~al.}(2002){Dennett-Thorpe}, {Scheuer}, {Laing},
  {Bridle}, {Pooley}, \& {Reich}}]{dennetthorpe02}
{Dennett-Thorpe}, J., {Scheuer}, P.~A.~G., {Laing}, R.~A., {et~al.} 2002,
  \mnras, 330, 609

\bibitem[{{Dey} {et~al.}(2019){Dey}, {Schlegel}, {Lang}, {Blum}, {Burleigh},
  {Fan}, {Findlay}, {Finkbeiner}, {Herrera}, {Juneau}, {Landriau}, {Levi},
  {McGreer}, {Meisner}, {Myers}, {Moustakas}, {Nugent}, {Patej}, {Schlafly},
  {Walker}, {Valdes}, {Weaver}, {Y{\`e}che}, {Zou}, {Zhou}, {Abareshi},
  {Abbott}, {Abolfathi}, {Aguilera}, {Alam}, {Allen}, {Alvarez}, {Annis},
  {Ansarinejad}, {Aubert}, {Beechert}, {Bell}, {BenZvi}, {Beutler}, {Bielby},
  {Bolton}, {Brice{\~n}o}, {Buckley-Geer}, {Butler}, {Calamida}, {Carlberg},
  {Carter}, {Casas}, {Castander}, {Choi}, {Comparat}, {Cukanovaite}, {Delubac},
  {DeVries}, {Dey}, {Dhungana}, {Dickinson}, {Ding}, {Donaldson}, {Duan},
  {Duckworth}, {Eftekharzadeh}, {Eisenstein}, {Etourneau}, {Fagrelius},
  {Farihi}, {Fitzpatrick}, {Font-Ribera}, {Fulmer}, {G{\"a}nsicke},
  {Gaztanaga}, {George}, {Gerdes}, {Gontcho}, {Gorgoni}, {Green}, {Guy},
  {Harmer}, {Hernandez}, {Honscheid}, {Huang}, {James}, {Jannuzi}, {Jiang},
  {Joyce}, {Karcher}, {Karkar}, {Kehoe}, {Kneib}, {Kueter-Young}, {Lan},
  {Lauer}, {Le Guillou}, {Le Van Suu}, {Lee}, {Lesser}, {Perreault Levasseur},
  {Li}, {Mann}, {Marshall}, {Mart{\'\i}nez-V{\'a}zquez}, {Martini}, {du Mas des
  Bourboux}, {McManus}, {Meier}, {M{\'e}nard}, {Metcalfe},
  {Mu{\~n}oz-Guti{\'e}rrez}, {Najita}, {Napier}, {Narayan}, {Newman}, {Nie},
  {Nord}, {Norman}, {Olsen}, {Paat}, {Palanque-Delabrouille}, {Peng},
  {Poppett}, {Poremba}, {Prakash}, {Rabinowitz}, {Raichoor}, {Rezaie},
  {Robertson}, {Roe}, {Ross}, {Ross}, {Rudnick}, {Safonova}, {Saha},
  {S{\'a}nchez}, {Savary}, {Schweiker}, {Scott}, {Seo}, {Shan}, {Silva},
  {Slepian}, {Soto}, {Sprayberry}, {Staten}, {Stillman}, {Stupak}, {Summers},
  {Sien Tie}, {Tirado}, {Vargas-Maga{\~n}a}, {Vivas}, {Wechsler}, {Williams},
  {Yang}, {Yang}, {Yapici}, {Zaritsky}, {Zenteno}, {Zhang}, {Zhang}, {Zhou}, \&
  {Zhou}}]{dey19}
{Dey}, A., {Schlegel}, D.~J., {Lang}, D., {et~al.} 2019, \aj, 157, 168

\bibitem[{{Duncan} {et~al.}(2021){Duncan}, {Kondapally}, {Brown}, {Bonato},
  {Best}, {R{\"o}ttgering}, {Bondi}, {Bowler}, {Cochrane}, {G{\"u}rkan},
  {Hardcastle}, {Jarvis}, {Kunert-Bajraszewska}, {Leslie}, {Ma{\l}ek},
  {Morabito}, {O'Sullivan}, {Prandoni}, {Sabater}, {Shimwell}, {Smith}, {Wang},
  {Wo{\l}owska}, \& {Tasse}}]{duncan21}
{Duncan}, K.~J., {Kondapally}, R., {Brown}, M.~J.~I., {et~al.} 2021, \aap, 648,
  A4

\bibitem[{{Ekers} {et~al.}(1978){Ekers}, {Fanti}, {Lari}, \& {Parma}}]{ekers78}
{Ekers}, R.~D., {Fanti}, R., {Lari}, C., \& {Parma}, P. 1978, \nat, 276, 588

\bibitem[{{Fanaroff} \& {Riley}(1974)}]{fanaroff74}
{Fanaroff}, B.~L. \& {Riley}, J.~M. 1974, \mnras, 167, 31P

\bibitem[{{Fragile} \& {Anninos}(2005)}]{fragile05}
{Fragile}, P.~C. \& {Anninos}, P. 2005, \apj, 623, 347

\bibitem[{{Fragner} \& {Nelson}(2010)}]{fragner10}
{Fragner}, M.~M. \& {Nelson}, R.~P. 2010, \aap, 511, A77

\bibitem[{{Garofalo} \& {Singh}(2019)}]{garofalo19}
{Garofalo}, D. \& {Singh}, C.~B. 2019, \apj, 871, 259

\bibitem[{{Giri} {et~al.}(2022){Giri}, {Vaidya}, {Rossi}, {Bodo}, {Mukherjee},
  \& {Mignone}}]{gourab22}
{Giri}, G., {Vaidya}, B., {Rossi}, P., {et~al.} 2022, \aap, 662, A5

\bibitem[{{Gopal-Krishna} \& {Wiita}(2000)}]{Gopal-Krishna+Wiita00}
{Gopal-Krishna} \& {Wiita}, P.~J. 2000, \aap, 363, 507

\bibitem[{{Gower} {et~al.}(1982){Gower}, {Gregory}, {Unruh}, \&
  {Hutchings}}]{gower82}
{Gower}, A.~C., {Gregory}, P.~C., {Unruh}, W.~G., \& {Hutchings}, J.~B. 1982,
  \apj, 262, 478

\bibitem[{{Hardcastle}(2018)}]{hardcastle18}
{Hardcastle}, M.~J. 2018, \mnras, 475, 2768

\bibitem[{{Hardcastle} {et~al.}(2019){Hardcastle}, {Croston}, {Shimwell},
  {Tasse}, {G{\"u}rkan}, {Morganti}, {Murgia}, {R{\"o}ttgering}, {van Weeren},
  \& {Williams}}]{hardcastle19}
{Hardcastle}, M.~J., {Croston}, J.~H., {Shimwell}, T.~W., {et~al.} 2019,
  \mnras, 488, 3416

\bibitem[{{Hardcastle} {et~al.}(2023){Hardcastle}, {Horton}, {Williams},
  {Duncan}, {Alegre}, {Barkus}, {Croston}, {Dickinson}, {Osinga},
  {R{\"o}ttgering}, {Sabater}, {Shimwell}, {Smith}, {Best}, {Botteon},
  {Br{\"u}ggen}, {Drabent}, {de Gasperin}, {G{\"u}rkan}, {Hajduk}, {Hale},
  {Hoeft}, {Jamrozy}, {Kunert-Bajraszewska}, {Kondapally}, {Magliocchetti},
  {Mahatma}, {Mostert}, {O'Sullivan}, {Pajdosz-{\'S}mierciak}, {Petley},
  {Pierce}, {Prandoni}, {Schwarz}, {Shulewski}, {Siewert}, {Stott}, {Tang},
  {Vaccari}, {Zheng}, {Bailey}, {Desbled}, {Goyal}, {Gonano}, {Hanset},
  {Kurtz}, {Lim}, {Mielle}, {Molloy}, {Roth}, {Terentev}, \&
  {Torres}}]{hardcastle23}
{Hardcastle}, M.~J., {Horton}, M.~A., {Williams}, W.~L., {et~al.} 2023, \aap,
  678, A151

\bibitem[{{Hardcastle} \& {Sakelliou}(2004)}]{hardcastle04}
{Hardcastle}, M.~J. \& {Sakelliou}, I. 2004, \mnras, 349, 560

\bibitem[{{Harwood} {et~al.}(2020){Harwood}, {Vernstrom}, \&
  {Stroe}}]{harwood20}
{Harwood}, J.~J., {Vernstrom}, T., \& {Stroe}, A. 2020, \mnras, 491, 803

\bibitem[{{Horton} {et~al.}(2020{\natexlab{a}}){Horton}, {Hardcastle}, {Read},
  \& {Krause}}]{horton20a}
{Horton}, M.~A., {Hardcastle}, M.~J., {Read}, S.~C., \& {Krause}, M. G.~H.
  2020{\natexlab{a}}, \mnras [\eprint[arXiv]{2002.04966}]

\bibitem[{{Horton} {et~al.}(2020{\natexlab{b}}){Horton}, {Krause}, \&
  {Hardcastle}}]{horton20b}
{Horton}, M.~A., {Krause}, M. G.~H., \& {Hardcastle}, M.~J. 2020{\natexlab{b}},
  \mnras, 499, 5765

\bibitem[{{Horton} {et~al.}(2023){Horton}, {Krause}, \&
  {Hardcastle}}]{horton23}
{Horton}, M.~A., {Krause}, M. G.~H., \& {Hardcastle}, M.~J. 2023, \mnras, 521,
  2593

\bibitem[{{Hunstead} {et~al.}(1984){Hunstead}, {Murdoch}, {Condon}, \&
  {Phillips}}]{hunstead84}
{Hunstead}, R.~W., {Murdoch}, H.~S., {Condon}, J.~J., \& {Phillips}, M.~M.
  1984, \mnras, 207, 55

\bibitem[{Hunter(2007)}]{Hunter07}
Hunter, J.~D. 2007, Computing In Science \& Engineering, 9, 90

\bibitem[{{Jones} \& {Owen}(1979)}]{Jones+Owen79}
{Jones}, T.~W. \& {Owen}, F.~N. 1979, \apj, 234, 818

\bibitem[{{Juneau} \& {Dey}(2022)}]{juneau22}
{Juneau}, S. \& {Dey}, A. 2022, The NOIRLab Mirror, 3, 10

\bibitem[{{Jurlin} {et~al.}(2021){Jurlin}, {Brienza}, {Morganti}, {Wadadekar},
  {Ishwara-Chandra}, {Maddox}, \& {Mahatma}}]{jurlin21}
{Jurlin}, N., {Brienza}, M., {Morganti}, R., {et~al.} 2021, \aap, 653, A110

\bibitem[{{Krause} {et~al.}(2019){Krause}, {Shabala}, {Hardcastle}, {Bicknell},
  {B{\"o}hringer}, {Chon}, {Nawaz}, {Sarzi}, \& {Wagner}}]{krause18}
{Krause}, M. G.~H., {Shabala}, S.~S., {Hardcastle}, M.~J., {et~al.} 2019,
  \mnras, 482, 240

\bibitem[{{Kumari} {et~al.}(2024){Kumari}, {Pal}, {Hardcastle}, \&
  {Horton}}]{shobha24}
{Kumari}, S., {Pal}, S., {Hardcastle}, M.~J., \& {Horton}, M.~A. 2024, \aap,
  689, A301

\bibitem[{{Leahy} \& {Williams}(1984)}]{leahy84}
{Leahy}, J.~P. \& {Williams}, A.~G. 1984, \mnras, 210, 929

\bibitem[{{Lilly} \& {Longair}(1984)}]{lilly84}
{Lilly}, S.~J. \& {Longair}, M.~S. 1984, \mnras, 211, 833

\bibitem[{{Mahatma}(2023)}]{mahatma23}
{Mahatma}, V.~H. 2023, Galaxies, 11, 74

\bibitem[{{Mahatma} {et~al.}(2019){Mahatma}, {Hardcastle}, {Williams}, {Best},
  {Croston}, {Duncan}, {Mingo}, {Morganti}, {Brienza}, {Cochrane},
  {G{\"u}rkan}, {Harwood}, {Jarvis}, {Jamrozy}, {Jurlin}, {Morabito},
  {R{\"o}ttgering}, {Sabater}, {Shimwell}, {Smith}, {Shulevski}, \&
  {Tasse}}]{mahatma19}
{Mahatma}, V.~H., {Hardcastle}, M.~J., {Williams}, W.~L., {et~al.} 2019,
  Astronomy and Astrophysics, 622, A13

\bibitem[{{Mahatma} {et~al.}(2018){Mahatma}, {Hardcastle}, {Williams},
  {Brienza}, {Br{\"u}ggen}, {Croston}, {Gurkan}, {Harwood},
  {Kunert-Bajraszewska}, {Morganti}, {R{\"o}ttgering}, {Shimwell}, \&
  {Tasse}}]{mahatma18}
{Mahatma}, V.~H., {Hardcastle}, M.~J., {Williams}, W.~L., {et~al.} 2018,
  \mnras, 475, 4557

\bibitem[{{Meisner} {et~al.}(2018){Meisner}, {Lang}, \& {Schlegel}}]{meisner18}
{Meisner}, A.~M., {Lang}, D., \& {Schlegel}, D.~J. 2018, Research Notes of the
  American Astronomical Society, 2, 1

\bibitem[{{Mentuch Cooper} {et~al.}(2023){Mentuch Cooper}, {Gebhardt}, {Davis},
  {Farrow}, {Liu}, {Zeimann}, {Ciardullo}, {Feldmeier}, {Drory}, {Jeong},
  {Benda}, {Bowman}, {Boylan-Kolchin}, {Ch{\'a}vez Ortiz}, {Debski}, {Dentler},
  {Fabricius}, {Farooq}, {Finkelstein}, {Gawiser}, {Gronwall}, {Hill}, {Hopp},
  {House}, {Janowiecki}, {Khoraminezhad}, {Kollatschny}, {Komatsu}, {Landriau},
  {Niemeyer}, {Lee}, {MacQueen}, {Mawatari}, {McKay}, {Ouchi}, {Poppe},
  {Saito}, {Schneider}, {Snigula}, {Thomas}, {Tuttle}, {Urrutia}, {Weiss},
  {Wisotzki}, {Zhang}, \& {HETDEX Collaboration}}]{mentuchcooper23}
{Mentuch Cooper}, E., {Gebhardt}, K., {Davis}, D., {et~al.} 2023, \apj, 943,
  177

\bibitem[{{Merritt} \& {Ekers}(2002)}]{merritt02}
{Merritt}, D. \& {Ekers}, R.~D. 2002, Science, 297, 1310

\bibitem[{{Miley}(1980)}]{miley80}
{Miley}, G. 1980, \araa, 18, 165

\bibitem[{{Mingarelli} {et~al.}(2025){Mingarelli}, {Blecha}, {Bogdanovi{\'c}},
  {Charisi}, {Chen}, {Escala}, {Goncharov}, {Graham}, {Komossa}, {McWilliams},
  {Schwartz}, \& {Zrake}}]{mingarelli25}
{Mingarelli}, C.~M.~F., {Blecha}, L., {Bogdanovi{\'c}}, T., {et~al.} 2025,
  Nature Astronomy [\eprint[arXiv]{2501.08956}]

\bibitem[{{Mingo} {et~al.}(2019){Mingo}, {Croston}, {Hardcastle}, {Best},
  {Duncan}, {Morganti}, {Rottgering}, {Sabater}, {Shimwell}, {Williams},
  {Brienza}, {Gurkan}, {Mahatma}, {Morabito}, {Prandoni}, {Bondi}, {Ineson}, \&
  {Mooney}}]{mingo19}
{Mingo}, B., {Croston}, J.~H., {Hardcastle}, M.~J., {et~al.} 2019, \mnras, 488,
  2701

\bibitem[{{Misra} {et~al.}(2023){Misra}, {Jamrozy}, \&
  {We{\.z}gowiec}}]{misra23}
{Misra}, A., {Jamrozy}, M., \& {We{\.z}gowiec}, M. 2023, \mnras, 523, 1648

\bibitem[{{Morabito} {et~al.}(2022){Morabito}, {Sweijen}, {Radcliffe}, {Best},
  {Kondapally}, {Bondi}, {Bonato}, {Duncan}, {Prandoni}, {Shimwell},
  {Williams}, {van Weeren}, {Conway}, \& {Calistro Rivera}}]{morabito22}
{Morabito}, L.~K., {Sweijen}, F., {Radcliffe}, J.~F., {et~al.} 2022, \mnras,
  515, 5758

\bibitem[{{Mostert} {et~al.}(2023){Mostert}, {Morganti}, {Brienza}, {Duncan},
  {Oei}, {R{\"o}ttgering}, {Alegre}, {Hardcastle}, \& {Jurlin}}]{mostert23}
{Mostert}, R. I.~J., {Morganti}, R., {Brienza}, M., {et~al.} 2023, \aap, 674,
  A208

\bibitem[{{Mostert} {et~al.}(2024){Mostert}, {Oei}, {Barkus}, {Alegre},
  {Hardcastle}, {Duncan}, {R{\"o}ttgering}, {van Weeren}, \&
  {Horton}}]{mostert24}
{Mostert}, R.~I.~J., {Oei}, M.~S.~S.~L., {Barkus}, B., {et~al.} 2024, \aap,
  691, A185

\bibitem[{{Nixon} \& {King}(2013)}]{nixon13}
{Nixon}, C. \& {King}, A. 2013, \apjl, 765, L7

\bibitem[{{Norris} {et~al.}(2021){Norris}, {Crawford}, \&
  {Macgregor}}]{norris21}
{Norris}, R.~P., {Crawford}, E., \& {Macgregor}, P. 2021, Galaxies, 9, 83

\bibitem[{{Oei} {et~al.}(2023){Oei}, {van Weeren}, {Gast}, {Botteon},
  {Hardcastle}, {Dabhade}, {Shimwell}, {R{\"o}ttgering}, \& {Drabent}}]{oei23}
{Oei}, M. S.~S.~L., {van Weeren}, R.~J., {Gast}, A. R.~D.~J.~G.~I.~B., {et~al.}
  2023, \aap, 672, A163

\bibitem[{{Owen} {et~al.}(1978){Owen}, {Burns}, \& {Rudnick}}]{Owen+78}
{Owen}, F.~N., {Burns}, J.~O., \& {Rudnick}, L. 1978, \apjl, 226, L119

\bibitem[{{Owen} \& {Laing}(1989)}]{owen+laing89}
{Owen}, F.~N. \& {Laing}, R.~A. 1989, \mnras, 238, 357

\bibitem[{{Parma} {et~al.}(2007){Parma}, {Murgia}, {de Ruiter}, {Fanti},
  {Mack}, \& {Govoni}}]{Parma+07}
{Parma}, P., {Murgia}, M., {de Ruiter}, H.~R., {et~al.} 2007, \aap, 470, 875

\bibitem[{{Pfeifle} {et~al.}(2024){Pfeifle}, {Weaver}, {Secrest}, {Rothberg},
  \& {Patton}}]{pfeifle24}
{Pfeifle}, R.~W., {Weaver}, K.~A., {Secrest}, N.~J., {Rothberg}, B., \&
  {Patton}, D.~R. 2024, arXiv e-prints, arXiv:2411.12799

\bibitem[{{Proctor}(2011)}]{proctor11}
{Proctor}, D.~D. 2011, \apjs, 194, 31

\bibitem[{{Reardon} {et~al.}(2023){Reardon}, {Zic}, {Shannon}, {Hobbs},
  {Bailes}, {Di Marco}, {Kapur}, {Rogers}, {Thrane}, {Askew}, {Bhat},
  {Cameron}, {Cury{\l}o}, {Coles}, {Dai}, {Goncharov}, {Kerr}, {Kulkarni},
  {Levin}, {Lower}, {Manchester}, {Mandow}, {Miles}, {Nathan}, {Os{\l}owski},
  {Russell}, {Spiewak}, {Zhang}, \& {Zhu}}]{reardon23}
{Reardon}, D.~J., {Zic}, A., {Shannon}, R.~M., {et~al.} 2023, \apjl, 951, L6

\bibitem[{{Riles}(2013)}]{riles13}
{Riles}, K. 2013, Progress in Particle and Nuclear Physics, 68, 1

\bibitem[{{Rudnick}(2021)}]{rudnick21}
{Rudnick}, L. 2021, Galaxies, 9, 85

\bibitem[{{Schoenmakers} {et~al.}(2000){Schoenmakers}, {de Bruyn},
  {R{\"o}ttgering}, {van der Laan}, \& {Kaiser}}]{schoenmakers00b}
{Schoenmakers}, A.~P., {de Bruyn}, A.~G., {R{\"o}ttgering}, H.~J.~A., {van der
  Laan}, H., \& {Kaiser}, C.~R. 2000, \mnras, 315, 371

\bibitem[{{Shimwell} {et~al.}(2022){Shimwell}, {Hardcastle}, {Tasse}, {Best},
  {R{\"o}ttgering}, {Williams}, {Botteon}, {Drabent}, {Mechev}, {Shulevski},
  {van Weeren}, {Bester}, {Br{\"u}ggen}, {Brunetti}, {Callingham}, {Chy{\.z}y},
  {Conway}, {Dijkema}, {Duncan}, {de Gasperin}, {Hale}, {Haverkorn}, {Hugo},
  {Jackson}, {Mevius}, {Miley}, {Morabito}, {Morganti}, {Offringa}, {Oonk},
  {Rafferty}, {Sabater}, {Smith}, {Schwarz}, {Smirnov}, {O'Sullivan},
  {Vedantham}, {White}, {Albert}, {Alegre}, {Asabere}, {Bacon}, {Bonafede},
  {Bonnassieux}, {Brienza}, {Bilicki}, {Bonato}, {Calistro Rivera}, {Cassano},
  {Cochrane}, {Croston}, {Cuciti}, {Dallacasa}, {Danezi}, {Dettmar}, {Di
  Gennaro}, {Edler}, {En{\ss}lin}, {Emig}, {Franzen}, {Garc{\'\i}a-Vergara},
  {Grange}, {G{\"u}rkan}, {Hajduk}, {Heald}, {Heesen}, {Hoang}, {Hoeft},
  {Horellou}, {Iacobelli}, {Jamrozy}, {Jeli{\'c}}, {Kondapally}, {Kukreti},
  {Kunert-Bajraszewska}, {Magliocchetti}, {Mahatma}, {Ma{\l}ek}, {Mandal},
  {Massaro}, {Meyer-Zhao}, {Mingo}, {Mostert}, {Nair}, {Nakoneczny},
  {Nikiel-Wroczy{\'n}ski}, {Orr{\'u}}, {Pajdosz-{\'S}mierciak}, {Pasini},
  {Prandoni}, {van Piggelen}, {Rajpurohit}, {Retana-Montenegro}, {Riseley},
  {Rowlinson}, {Saxena}, {Schrijvers}, {Sweijen}, {Siewert}, {Timmerman},
  {Vaccari}, {Vink}, {West}, {Wo{\l}owska}, {Zhang}, \& {Zheng}}]{shimwell22}
{Shimwell}, T.~W., {Hardcastle}, M.~J., {Tasse}, C., {et~al.} 2022, \aap, 659,
  A1

\bibitem[{{Shimwell} {et~al.}(2017){Shimwell}, {R{\"o}ttgering}, {Best},
  {Williams}, {Dijkema}, {de Gasperin}, {Hardcastle}, {Heald}, {Hoang},
  {Horneffer}, {Intema}, {Mahony}, {Mandal}, {Mechev}, {Morabito}, {Oonk},
  {Rafferty}, {Retana-Montenegro}, {Sabater}, {Tasse}, {van Weeren},
  {Br{\"u}ggen}, {Brunetti}, {Chy{\.z}y}, {Conway}, {Haverkorn}, {Jackson},
  {Jarvis}, {McKean}, {Miley}, {Morganti}, {White}, {Wise}, {van Bemmel},
  {Beck}, {Brienza}, {Bonafede}, {Calistro Rivera}, {Cassano}, {Clarke},
  {Cseh}, {Deller}, {Drabent}, {van Driel}, {Engels}, {Falcke}, {Ferrari},
  {Fr{\"o}hlich}, {Garrett}, {Harwood}, {Heesen}, {Hoeft}, {Horellou},
  {Israel}, {Kapi{\'n}ska}, {Kunert-Bajraszewska}, {McKay}, {Mohan},
  {Orr{\'u}}, {Pizzo}, {Prandoni}, {Schwarz}, {Shulevski}, {Sipior}, {Smith},
  {Sridhar}, {Steinmetz}, {Stroe}, {Varenius}, {van der Werf}, {Zensus}, \&
  {Zwart}}]{shimwell17}
{Shimwell}, T.~W., {R{\"o}ttgering}, H.~J.~A., {Best}, P.~N., {et~al.} 2017,
  \aap, 598, A104

\bibitem[{{Shimwell} {et~al.}(2019){Shimwell}, {Tasse}, {Hardcastle}, {Mechev},
  {Williams}, {Best}, {R{\"o}ttgering}, {Callingham}, {Dijkema}, {de Gasperin},
  {Hoang}, {Hugo}, {Mirmont}, {Oonk}, {Prandoni}, {Rafferty}, {Sabater},
  {Smirnov}, {van Weeren}, {White}, {Atemkeng}, {Bester}, {Bonnassieux},
  {Br{\"u}ggen}, {Brunetti}, {Chy{\.z}y}, {Cochrane}, {Conway}, {Croston},
  {Danezi}, {Duncan}, {Haverkorn}, {Heald}, {Iacobelli}, {Intema}, {Jackson},
  {Jamrozy}, {Jarvis}, {Lakhoo}, {Mevius}, {Miley}, {Morabito}, {Morganti},
  {Nisbet}, {Orr{\'u}}, {Perkins}, {Pizzo}, {Schrijvers}, {Smith}, {Vermeulen},
  {Wise}, {Alegre}, {Bacon}, {van Bemmel}, {Beswick}, {Bonafede}, {Botteon},
  {Bourke}, {Brienza}, {Calistro Rivera}, {Cassano}, {Clarke}, {Conselice},
  {Dettmar}, {Drabent}, {Dumba}, {Emig}, {En{\ss}lin}, {Ferrari}, {Garrett},
  {G{\'e}nova-Santos}, {Goyal}, {G{\"u}rkan}, {Hale}, {Harwood}, {Heesen},
  {Hoeft}, {Horellou}, {Jackson}, {Kokotanekov}, {Kondapally},
  {Kunert-Bajraszewska}, {Mahatma}, {Mahony}, {Mandal}, {McKean}, {Merloni},
  {Mingo}, {Miskolczi}, {Mooney}, {Nikiel-Wroczy{\'n}ski}, {O'Sullivan},
  {Quinn}, {Reich}, {Roskowi{\'n}ski}, {Rowlinson}, {Savini}, {Saxena},
  {Schwarz}, {Shulevski}, {Sridhar}, {Stacey}, {Urquhart}, {van der Wiel},
  {Varenius}, {Webster}, \& {Wilber}}]{shimwell19}
{Shimwell}, T.~W., {Tasse}, C., {Hardcastle}, M.~J., {et~al.} 2019, Astronomy
  and Astrophysics, 622, A1

\bibitem[{{Slee} {et~al.}(2001){Slee}, {Roy}, {Murgia}, {Andernach}, \&
  {Ehle}}]{slee01}
{Slee}, O.~B., {Roy}, A.~L., {Murgia}, M., {Andernach}, H., \& {Ehle}, M. 2001,
  \aj, 122, 1172

\bibitem[{{Smith} {et~al.}(2016){Smith}, {Best}, {Duncan}, {Hatch}, {Jarvis},
  {R{\"o}ttgering}, {Simpson}, {Stott}, {Cochrane}, {Coppin}, {Dannerbauer},
  {Davis}, {Geach}, {Hale}, {Hardcastle}, {Hatfield}, {Houghton}, {Maddox},
  {McGee}, {Morabito}, {Nisbet}, {Pandey-Pommier}, {Prandoni}, {Saxena},
  {Shimwell}, {Tarr}, {van Bemmel}, {Verma}, {White}, \& {Williams}}]{smith16}
{Smith}, D.~J.~B., {Best}, P.~N., {Duncan}, K.~J., {et~al.} 2016, in SF2A-2016:
  Proceedings of the Annual meeting of the French Society of Astronomy and
  Astrophysics, ed. C.~{Reyl{\'e}}, J.~{Richard}, L.~{Cambr{\'e}sy},
  M.~{Deleuil}, E.~{P{\'e}contal}, L.~{Tresse}, \& I.~{Vauglin}, 271--280

\bibitem[{{Steinle} {et~al.}(2024){Steinle}, {Gerosa}, \& {Krause}}]{steinle24}
{Steinle}, N., {Gerosa}, D., \& {Krause}, M. G.~H. 2024, \prd, 110, 123034

\bibitem[{{Stone} \& {Loeb}(2012)}]{stone12}
{Stone}, N. \& {Loeb}, A. 2012, \prl, 108, 061302

\bibitem[{{Taylor}(2005)}]{Taylor05}
{Taylor}, M.~B. 2005, in Astronomical Society of the Pacific Conference Series,
  Vol. 347, Astronomical Data Analysis Software and Systems XIV, ed.
  P.~{Shopbell}, M.~{Britton}, \& R.~{Ebert}, 29

\bibitem[{{van Weeren} {et~al.}(2019){van Weeren}, {de Gasperin}, {Akamatsu},
  {Br{\"u}ggen}, {Feretti}, {Kang}, {Stroe}, \& {Zandanel}}]{vanweeren19}
{van Weeren}, R.~J., {de Gasperin}, F., {Akamatsu}, H., {et~al.} 2019, \ssr,
  215, 16

\bibitem[{{Williams} {et~al.}(2019){Williams}, {Hardcastle}, {Best}, {Sabater},
  {Croston}, {Duncan}, {Shimwell}, {R{\"o}ttgering}, {Nisbet}, {G{\"u}rkan},
  {Alegre}, {Cochrane}, {Goyal}, {Hale}, {Jackson}, {Jamrozy}, {Kondapally},
  {Kunert-Bajraszewska}, {Mahatma}, {Mingo}, {Morabito}, {Prandoni},
  {Roskowinski}, {Shulevski}, {Smith}, {Tasse}, {Urquhart}, {Webster}, {White},
  {Beswick}, {Callingham}, {Chy{\.z}y}, {de Gasperin}, {Harwood}, {Hoeft},
  {Iacobelli}, {McKean}, {Mechev}, {Miley}, {Schwarz}, \& {van
  Weeren}}]{williams19}
{Williams}, W.~L., {Hardcastle}, M.~J., {Best}, P.~N., {et~al.} 2019, \aap,
  622, A2

\bibitem[{{Yang} {et~al.}(2019){Yang}, {Joshi}, {Gopal-Krishna}, {An}, {Ho},
  {Wiita}, {Liu}, {Yang}, {Wang}, {Wu}, \& {Yang}}]{yang19}
{Yang}, X., {Joshi}, R., {Gopal-Krishna}, {et~al.} 2019, \apjs, 245, 17

\bibitem[{{Zirbel}(1996)}]{zirbel96}
{Zirbel}, E.~L. 1996, \apj, 473, 713

\end{thebibliography}
\clearpage

\end{document}